\documentclass[prl,aps,twocolumn,floats,nofootinbib]{revtex4}
\usepackage{amsmath,amssymb,graphicx,psfrag}
\usepackage{graphicx}
\usepackage{dcolumn}
\usepackage{bm}
\usepackage{subfigure}

\begin{document}
\renewcommand{\ni}{{\noindent}}
\newcommand{\dprime}{{\prime\prime}}
\newcommand{\be}{\begin{equation}}
\newcommand{\ee}{\end{equation}}
\newcommand{\bea}{\begin{eqnarray}} 
\newcommand{\eea}{\end{eqnarray}}
\newcommand{\la}{\langle}
\newcommand{\ra}{\rangle} 
\newcommand{\dg}{\dagger}
\newcommand\lbs{\left[}
\newcommand\rbs{\right]}
\newcommand\lbr{\left(}
\newcommand\rbr{\right)}
\newcommand\f{\frac}
\newcommand\e{\epsilon}
\newcommand\ua{\uparrow}
\newcommand\da{\downarrow}
\title{Gutzwiller projection for exclusion of holes:  Application to strongly correlated Ionic Hubbard Model and Binary Alloys} 

\author{Anwesha Chattopadhyay and Arti Garg} 
\affiliation{Condensed Matter Physics Division, Saha Institute of Nuclear Physics, HBNI, 1/AF Bidhannagar, Kolkata 700 064, India}
\email{anwesha.chattopadhyay@saha.ac.in}
\email{arti.garg@saha.ac.in}
\vspace{0.2cm}

\begin{abstract}
We consider strongly correlated limit of variants of the Hubbard model (HM) in which on parts of the system it is energetically favourable to project out doublons from the low energy Hilbert space while on other sites of the system it is favourable to project out holes while still allowing for doublons. As an effect the low energy Hilbert space itself varies with sites of the system. Though the formalism is well developed for the case of doublon projection in the literature, case of hole projection has not been explored in detail so far. We derive basic framework by defining creation and annihilation operators for electrons in a restricted Hilbert space where holes are projected out but which still allows for doublons. We generalise the idea of Gutzwiller approximation for case of hole projection which has been done in literature for the case of doublon projection. To be specific, we provide detailed analysis of strongly correlated limit of the ionic Hubbard model (IHM) which has a staggered potential $\Delta$ on two sublattices of a bipartite lattice and the correlated binary alloys which have binary disorder $\pm V/2$ randomly distributed on sites of the lattice. In both the cases, for $\Delta \sim U \gg t$ and for $V\sim U \gg t$, where $U$ is the Hubbard energy cost for having a doublon at a site, there are sites on which doublons are allowed while holes are the maximum energy states. We do a systematic generalization of similarity transformation for both these cases and obtain the effective low energy Hamiltonian. We further derive Gutzwiller approximation factors which provide renormalization of various terms in the effective low energy Hamiltonian due to the Gutzwiller projection operators, excluding holes on some sites and doublons on the remaining sites.
\end{abstract}
\maketitle

\section{Introduction}

Strongly correlated systems are of immense interest and importance in condensed matter physics. Strong e-e interactions leads to many interesting phases like high-$T_c$ superconductivity, anti-ferromagnetically ordered phase and Mott insulator. Hubbard model is a paradigmatic model in strongly correlated electron systems with two simple ingredients, namely, hopping of electrons ($\sim t$) and onsite Coulomb interaction($\sim U$). In the limit of large $U$ and finite hole doping, doublons are energetically unfavourable and needs to be projected out from the low energy Hilbert space. A regular similarity transformation which projects out double occupancies, gives the effective low energy Hamiltonian which is known as the $t-J$ model~\cite{Fazekas} and captures many aspects of the physics of high $T_c$  superconducting cuprates~\cite{high_Tc}. 

  The $t-J$ model is defined in the projected Hilbert space and since Wick's theorem does not work for the fermionic operators in the projected Hilbert space, standard many-body physics tools of calculating various order Feynman diagrams for the self-energy~\cite{Fetter} can not be used to solve this model. One needs to solve the Schwinger equation of motion for the Green's function of projected electrons ~\cite{Shastry} and do a systematic perturbation theory in some parameter that controls double occupancy. Numerically $t-J$ model can be studied using variational Monte Carlo method~\cite{VMC} where one starts with a variational wavefunction and then carry out doublon projection from each site explicitly . But because of the computational complexity, another alternative analytical tool is most commonly used in the community which is an approximate way of implementing the Gutzwiller projection (elimination of double occupancies) and is known as Gutzwiller approximation. Gutzwiller approximation, as first introduced by Gutzwiller~\cite{Gutzwiller}, was improved and investigated later by several others~\cite{GA} mainly in context of hole-doped $t-J$ model. Under this approximation, the expectation values in the projected state is related to that in the un-projected state by a classical statistical weight factor know as the Gutzwiller factor that accounts for doublon exclusion. As an effect various terms in the Hamiltonian get renormalised by the Gutzwiller factors and the renormalised Hamiltonian can be studied in the unprojected basis. 

Though the Gutzwiller projection for exclusion of doublons has been explored in detail in the literature, Gutzwiller projection of holes from the low energy Hilbert space and its implementation in renormalizing the couplings in the effective low energy Hamiltonian at the level of Gutzwiller approximation is still completely unexplored. There are models, like electron doped $t-J$ model, where in the low energy Hilbert space one has to allow for doublons and holes have to be excluded. But in this situation it is not really essential to use the formalism of Gutzwiller projection for holes as one can simply do particle-hole transformation and map the model to hole-doped $t-J$ model where the low energy Hilbert space allows for holes excluding doublons. Hence probably the formalism of Gutzwiller projection of holes has not been explored yet. But there are situations where Gutzwiller projection of holes become crucial to carry out e.g. in a model where on some of the sites it is energetically favourable to do hole projection while on some other sites doublon projection is required. With this motivation, we provide basic formalism for Gutzwiller projection of holes and calculate the Gutzwiller factors for implementing this projection approximately by renormalizing the couplings in the low energy Hamiltonian for a couple of such models.  

In this work we provide a general formalism for studying variants of the strongly correlated Hubbard model with inhomogeneous onsite potential terms of the same order as $U$ or larger than that. Due to competing effects of onsite potential and $U$, there are sites at which holes are the maximum energy states (rather than doublons) and should be projected out from the low energy Hilbert space. We do a systematic extension of the similarity transformation in which the similarity operator itself varies from bond to bond depending upon whether both sites of the bond have doublons projected low energy Hilbert space dominated by large $U$ physics, or both have hole projected low energy Hilbert space or one of the site on the bond has a hole projected and the other site has a doublon projected low energy Hilbert space. 
We further calculate generalised Gutzwiller approximation factors for various terms in the low energy effective Hamiltonian which are also bond dependent. Gutzwiller factors for bonds where one site requires hole projection and the other has doublon projection or where both the sites have hole projecton have not been calculated in the literature earlier and in this work we derive them under the assumption that spin resolved densities before and after the projection remain the same. 

To be specific, we provide details of the formalism for two well studied models, namely, ionic Hubbard model (IHM) and correlated binary alloys represented by the Hubbard model in the presence of binary disorder. IHM is an interesting extension of the Hubbard model with a staggered onsite potential $\Delta$ added onto it. IHM has been studied in various dimensions by a variety of numerical and analytical tools. In one-dimension~\cite{1d_IHM}, it has been shown to have a spontaneously dimerised phase, in the intermediate coupling regime, which separates the weakly coupled band insulator from the strong coupling Mott insulator. In higher dimensions ($d > 1$), this model has been studied mainly using dynamical mean field theory (DMFT)~\cite{Jabben,AG1,hartmann,kampf,AG2,rajdeep,Kim,Soumen}, determinantal quantum Monte carlo~\cite{qmc_ihm1,qmc_ihm2}, cluster DMFT~\cite{cdmft} and coherent potential approximation~\cite{cpa}. Though the solution of DMFT self consistent equations in the paramagnetic (PM) sector at half filling at zero temperature shows an intervening metallic phase~\cite{AG1}, in the spin asymmetric sector, the transition from paramagnetic band insulator (PM BI) to anti-ferromagnetic (AFM) insulator preempts the formation of a para-metallic phase~\cite{kampf,cdmft}. In a recent work coauthored by one of us, it was shown that upon doping the IHM one gets a broad ferrimangetic half-metal phase~\cite{AG2}  sandwiched between a PM BI and a PM metal.  IHM has also been realised in optical lattices~\cite{expt_IHM} on honeycomb structure. 

Most of these earlier works on IHM are in the limit of weak to intermediate $U/t$ except ~\cite{Soumen,rajdeep} where strongly correlated limit of IHM has been studied for $\Delta \le U$ within DMFT. Recently~\cite{Rajdeep2} $\Delta \sim U\gg t$ limit of IHM has been studied using slave-boson mean field theory. Gutzwiller approximation method has been used for studying IHM ~\cite{Li} but in the limit of large U (not extreme correlation limit) where double occupancies are not fully prohibited. To the best of our knowledge, the Gutzwiller approximation formalism for this model has not been developed in the limit $\Delta\sim U\gg t$ which we present in this work. In the limit of large U and $\Delta$ ($U\sim \Delta$), holes are energetically expensive in the sublattice  where staggered potential is $-\Delta/2$ (say, sublattice A) and double occupancies are expensive in the  sublattice having potential $\Delta/2$(say B).  Therefore holes are projected out from A sublattice and doublons from B sublattice, which gives us the low energy effective Hamiltonian.

The second model for which we provide details of the formalism is the model of correlated binary alloys described by the Hubbard model in the presence of binary disorder potential. In all correlated electron systems, disorder is almost inevitable due to various intrinsic and extrinsic sources of impurities. In high $T_c$ cuprates, it is the doping of parent compound (e.g. with oxygen) which results in random onsite potential along with introducing holes~\cite{Pan}. Another type of common disorder is binary disorder which is for example realised in  disulfides ($Co_{1-x}Fe_xS_2$ and $Ni_{1-x}Co_xS_2$)~\cite{disulfides} in which two different transition metal ions are located at random positions, creating two different atomic levels for the correlated d-electrons. Binary disorder along with interactions among basic degrees of freedom has also been realised in optical lattice experiments~\cite{expt_binary}. Hence it becomes crucial to study interplay of disorder and interactions in order to understand many interesting properties of these systems. 

In correlated binary alloy model, onsite potential can be $\pm V/2$ at any site of the lattice randomly. The physics of this model has been explored for intermediate to strong coupling regime mainly using DMFT~\cite{Hofstetter1,Byczuk1,Byczuk2,Hassan}.  But the limit of large onsite repulsion as well as strong disorder potential $U \sim V \gg t$, where holes are projected out from sites having potential $-V/2$ (A) sites and double occupancies are projected out from sites having potential $V/2$ (B) sites, has not been explored so far. Though this model has similarity with the IHM mentioned above, but the intrinsic randomness associated with the binary disorder model makes the effective low energy Hamiltonian different from the case of IHM. Interplay of disorder and interaction in this model may lead to very different physics like many-body localization~\cite{MBL}. 

The rest of the paper is structured as follows. First we provide basic formalism for hole projection by defining electron creation and annihilation operators in the hole projected Hilbert space. We enlist probabilities of various allowed configurations in the hole projected Hilbert space and calculate the Gutzwiller approximation factors for hopping processes.  
In the next section, we have derived the effective low energy Hamiltonian for the IHM in the limit of $U \sim \Delta \gg t$ and calculated the corresponding Gutzwiller approximation factors for various terms in the Hamiltonian. Followed by this we  have described the similarity transformation and Gutzwiller approximation for correlated binary alloy in the limit of strong interactions and strong disorder. 
At the end, we also touch upon the case of fully random disorder and randomly distributed attractive impurities in the limit of both interaction and disorder strength being much larger than the hopping amplitude.

\section{Basic formalism for hole projection}
Though the formalism of Gutzwiller projection is well developed for the case of doublon projection in the literature, case of hole projection has not been explored in detail so far. In this section we derive basic framework by defining new creation and annihilation operators for electrons in a restricted Hilbert space where holes are projected out but which still allows for doublons.

\begin{figure}[h]
    \centering
    \includegraphics[width=5.5cm]{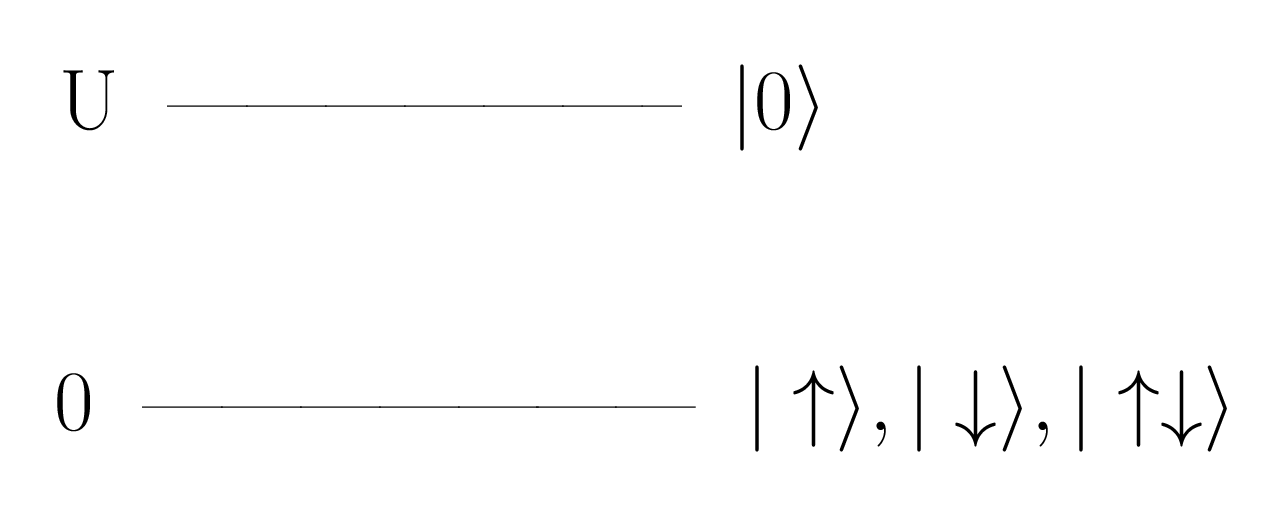}
    \caption{Separation in the energy scales of a hole and other states.}
  \label{level1}
\end{figure}

For a system of spin-1/2 fermions, at each site there are four possibilities, namely, $|\ua\ra, |\da\ra, |\ua\da\ra$ and $|0\ra$. Consider a model in which energy cost of having $|0\ra $ is much more than the other three states e.g., shown in Fig.~\ref{level1}. It may also happen that due to some other constraints e.g. to achieve certain density of particles in the system, one has to retain doublons in the low energy Hilbert space (though the energy cost for doublons  might be close to that of holes) and exclude holes. In these situations, the effective creation and annihilation operators for fermions in the low energy Hilbert space need to be modified.

The simplest way to see through this is following. Normal electron creation operator can be expressed in terms of local Hubbard operators:
\begin{equation}
c_{\sigma}^{\dagger}=X^{\sigma\leftarrow 0}+\eta(\sigma)X^{d \leftarrow \bar{\sigma}} 
\label{defn_c}                  
\end{equation}
where $\sigma$ can be $\uparrow$ or $\downarrow$ and $d$ represents a double occupancy and  $\eta(\uparrow)=1$ and $\eta(\downarrow)=-1$. Here we have used the local Hubbard operators defined as $ X^{b\leftarrow a}=|b\rangle\langle a|$. 
\begin{figure}[h]
    \centering
    \includegraphics[width=6cm]{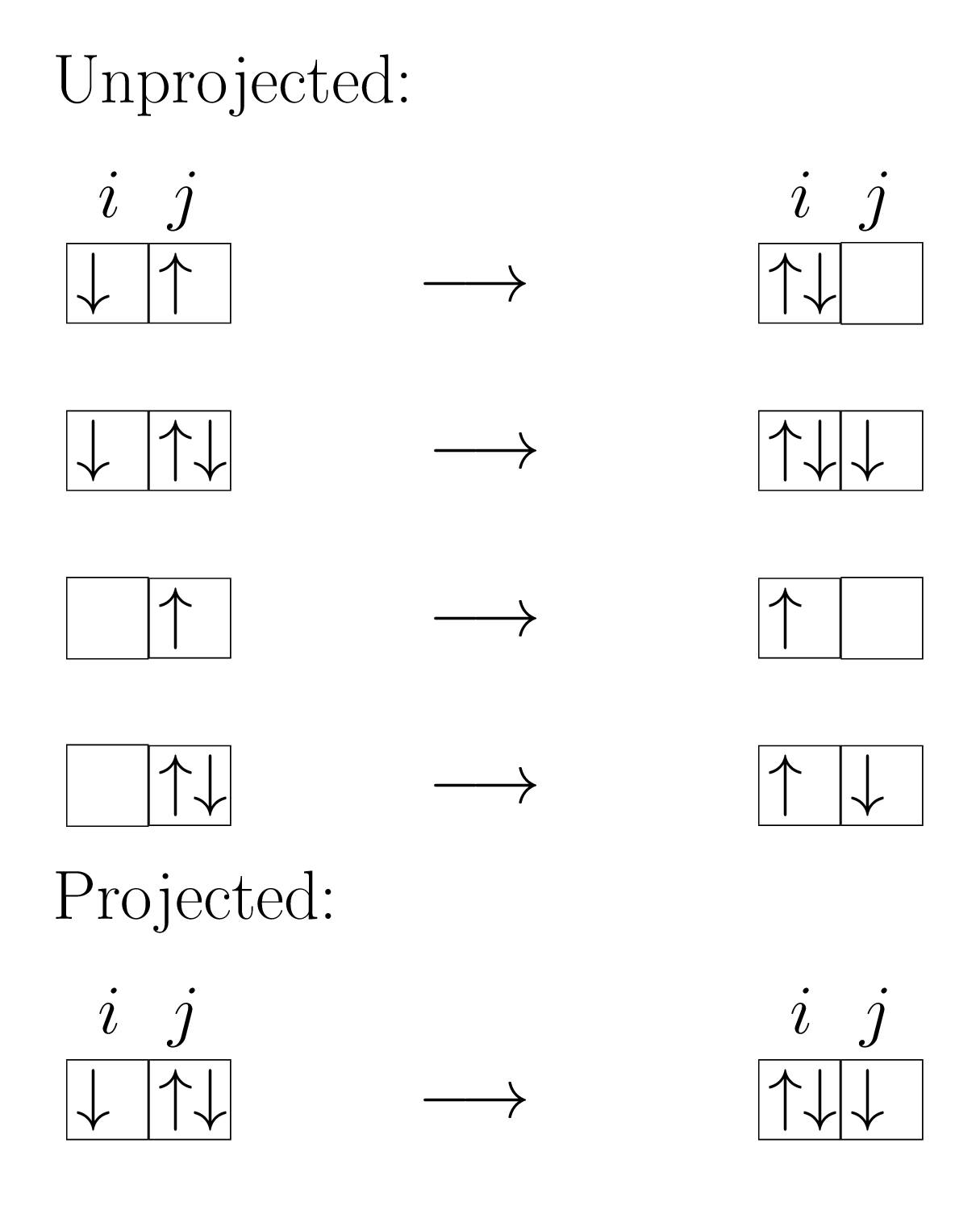}
    \caption{Top: Possible nearest neighbour hopping process in full Hilbert space. Bottom panel shows allowed hopping process in reduced Hilbert space from which hole has been projected out.}
  \label{hopp}
\end{figure}

This means one can create a particle either starting from a hole or by annihilating one particle from a double occupancy. Since, in the present case holes are projected out from the low energy subspace, one can not create a particle starting from a hole rather we can create a particle only by annihilating one particle from a doublon. Therefore, projected electron creation operator, which we denote by $\tilde{c}^\dagger_\sigma$, is

\begin{equation}
 \tilde{c}_{\sigma}^{\dagger}=\eta (\sigma)X^{d \leftarrow \bar{\sigma}}=c_{\sigma}^{\dagger}n_{\bar{\sigma}}
\label{defn_ctilde} 
\end{equation}
with $\eta(\ua)=1$ and $\eta(\da)=-1$. Note that $\tilde{c}_\sigma$ does not satisfy standard Lie algebra of fermions but $\{\tilde{c}_\sigma,\tilde{c}^\dagger_{\sigma}\} = n_{\bar{\sigma}}$. 
The corresponding number operator in this reduced Hilbert space is $\tilde{n}_\sigma = n_\sigma n_{\bar{\sigma}}$. Various Hubbard operators in form of fermionic operator in hole projected Hilbert space are given as $ X^{\sigma \leftarrow \sigma}= \tilde{c}_{\bar{\sigma}}\tilde{c}_{\bar{\sigma}}^{\dagger}$,  $X^{\sigma \leftarrow\bar{\sigma}}=-\tilde{c}_{\bar{\sigma}}\tilde{c_{\sigma}}^{\dagger}$ and $ X^{d \leftarrow d}=\tilde{c}_{\uparrow}^{\dagger}\tilde{c}_{\uparrow}=\tilde{c}_{\downarrow}^{\dagger}\tilde{c}_{\downarrow}$.
From the completeness relation of $X$ operators in hole projected Hilbert space we get
\bea 
X^{\ua\leftarrow \ua}+X^{\da\leftarrow \da}+X^{d\leftarrow d} = \mathcal{I} \nonumber \\
 n_\ua(1-n_\da)+n_\da(1-n_\ua)+n_\ua n_\da = \mathcal{I} \nonumber \\
 n_\ua n_\da = n-\mathcal{I}
\label{X_op}
\eea

Let us consider hopping of a particle to its nearest neighbour site in this reduced Hilbert space. In the full Hilbert space, which does not have constraint of hole projection, there are four possible nearest neighbour hopping processes as shown in the top panel of Fig.~\ref{hopp}.
But only allowed hopping processes in the low energy Hilbert space of hole projected system are those which do not have a hole in the initial state and in which no hole is created in the final state as well. This leaves for only one process in which there is a doublon at site $j$, and a spin $|\sigma\ra $ at site $i$. Then a $\bar{\sigma}$ hopes from site $j$ to $i$ resulting in a single occupancy at site $j$ and a doublon at site $i$ as shown in the bottom panel of Fig.~\ref{hopp}. Thus effectively only hopping of doublons takes place in the projected space resulting in an overall suppression of the hopping process. 

The corresponding operator for this hopping process is 
\bea
H_{hopp} = -t\sum_{<i,j>,\sigma} X_i^{d\leftarrow{\bar{\sigma}}}X_j^{\bar{\sigma}\leftarrow d} +h.c. \nonumber \\
= -t\sum_{<i,j>,\sigma} \tilde{c}_{i\sigma}^\dagger \tilde{c}_{j\sigma} + h.c.
\eea
 which is equivalently written in terms of normal fermionic operators as
\bea
 H_{hopp}=-t\sum_{<i,j>,\sigma} c_{i\sigma}^\dagger n_{i\bar{\sigma}}n_{j\bar{\sigma}}c_{j\sigma} + h.c. \nonumber \\
 = -\mathcal{P}_h(t\sum_{<i,j>,\sigma}c_{i\sigma}^{\dagger}c_{j\sigma}+h.c.)\mathcal{P}_h
\eea
Here $\mathcal{P}_h$ stands for the Gutzwiller projection operator for hole projection defined as $\mathcal{P}_{h}=\prod_{i}(1-(1-n_{i\uparrow})(1-n_{i\downarrow}))$. We now generalise the concept of Gutzwiller approximation for hole projected Hilbert space. The expectation value of the hopping process in the hole-projected Hilbert space can be obtained through Gutzwiller approximation by renormalizing the hopping term in the unprojected basis by a Gutzwiller factor which takes into account of the physics of projection approximately.
The Gutzwiller renormalization factor then is defined as the ratio of the expectation value of an operator $O$ in the projected basis to that in the unprojected basis:

\begin{equation}
    g=\dfrac{\langle \psi | \mathcal{P}_{h}  O \mathcal{P}_{h} | \psi \rangle}{\langle \psi | O | \psi \rangle}
\end{equation}
where, $\psi$ is the unprojected state.

The Gutzwiller renormalization factors are determined by the ratios of the probabilities of the corresponding physical processes in the projected and unprojected basis. 
Enlisted in Table.1 are the probabilities of states in unprojected and hole projected spaces where the spin resolved unprojected and projected densities have been taken to be equal.
{
\begin{center}
 \begin{tabular}{||c | c | c||} 
 \hline
 States & Unprojected  & Projected\\ [0.5ex] 
 \hline\hline
$ |\uparrow\rangle$ & ${\bf{n}}_{\uparrow}(1 - {\bf{n}}_{\downarrow})$ &  $(1 - {\bf{n}}_{\downarrow})$ \\ \hline
 $|\downarrow\rangle$ & ${\bf{n}}_{\downarrow}(1 - {\bf{n}}_{\uparrow})$ &  $(1 - {\bf{n}}_{\uparrow})$ \\ 
 \hline
$ |\uparrow\downarrow\rangle $ & ${\bf{n}}_{\uparrow}{\bf{n}}_{\downarrow}$ & $({\bf{n}} - 1)$ \\ 
\hline
$|0\rangle$ & $(1 - {\bf{n}}_{\uparrow})(1 - {\bf{n}}_{\downarrow})$ & $0$ \\ [1ex] 
 \hline
\end{tabular}
\end{center}
}
\normalsize

Table 1. Probabilities of different states in terms of e densities in unprojected and  hole projected basis.
\\
\vskip0.3cm
Here $\bf{n_\sigma}$ is the density of electron with spin $\sigma$. Consistently everywhere we use $\bf{n}$ for density and $n$ for the corresponding number operator. 

 The probability of hopping of an $\ua$ spin electron in the unprojected basis is $(1-{\bf{n}}_{i\uparrow}){\bf{n}}_{j\uparrow}{\bf{n}}_{i \uparrow}(1-{\bf{n}}_{j\uparrow})$. In the hole projected basis, the corresponding probability is $({\bf{n}}_{j}-1)({\bf{n}}_{i}-1)(1-{\bf{n}}_{i\uparrow})(1-{\bf{n}}_{j\uparrow})$. Therefore, the Gutzwiller factor for hopping process comes out to be 
\be
g_{t\uparrow}=\sqrt{\dfrac{({\bf{n}}_i-1)({\bf{n}}_j-1)}{{\bf{n}}_{i\uparrow}{\bf{n}}_{j\uparrow}}}
\label{gt}
\ee

With this set up for the hole projected Hilbert space, we describe strongly correlated limit of IHM and binary alloys.

\section{Strongly Correlated Limit of Ionic Hubbard Model}

IHM has tight-binding electrons on a bipartite lattice (sub-lattices A and B) described by the Hamiltonian
\[
H=-t\sum_{i\in A,j\in B,\sigma} [~c^{\dagger}_{i\sigma}c_{j\sigma}
+h.c~]- \f{\Delta}{2} \sum_{i\in A}n_{i} +\f{\Delta}{2}\sum_{i \in B} n_{i} \]
\be
\mbox{~~~~~~~~~~~~~~~~~~~~~~~~~~~}+U\sum_{i}n_{i\ua}n_{i\da}-\mu\sum_{i}n_{i} 
\label {model}
\ee
Here $t$ is the nearest neighbor hopping, $U$ the Hubbard repulsion and $\Delta$ a one-body
staggered potential which doubles the unit cell. The chemical potential is  $\mu=U/2$ for the
average occupancy per site to be one, that is, $\left(\langle n_A \rangle + \langle n_B \rangle \right)/2=1$, corresponding to ``half-filling''. 

Let us consider the t=0 limit of this model in the regime $U\sim \Delta$. On A sublattice, single occupancies have energy $-\bigg(\dfrac{\Delta}{2}+\dfrac{U}{2}\bigg)\sim -\Delta$, hole has $0$ energy and doublon has energy $-\Delta$. So, among the four choices of occupancy, a hole on A is the highest energy state and should be projected out from the low energy Hilbert space. On the other hand, on B sublattice, single occupancies cost $\bigg(\dfrac{\Delta}{2}-\dfrac{U}{2}\bigg)\sim 0$ energy, holes also cost $0$ energy while doublon cost energy $\Delta \sim U$ and therefore, on B sublattice, doublons should be projected out from the low energy Hilbert space. 

\subsection{Low Energy Hamiltonian in the limit $U\sim \Delta>>t$}
\begin{figure}[h]  
    \centering
    \includegraphics[width=8cm]{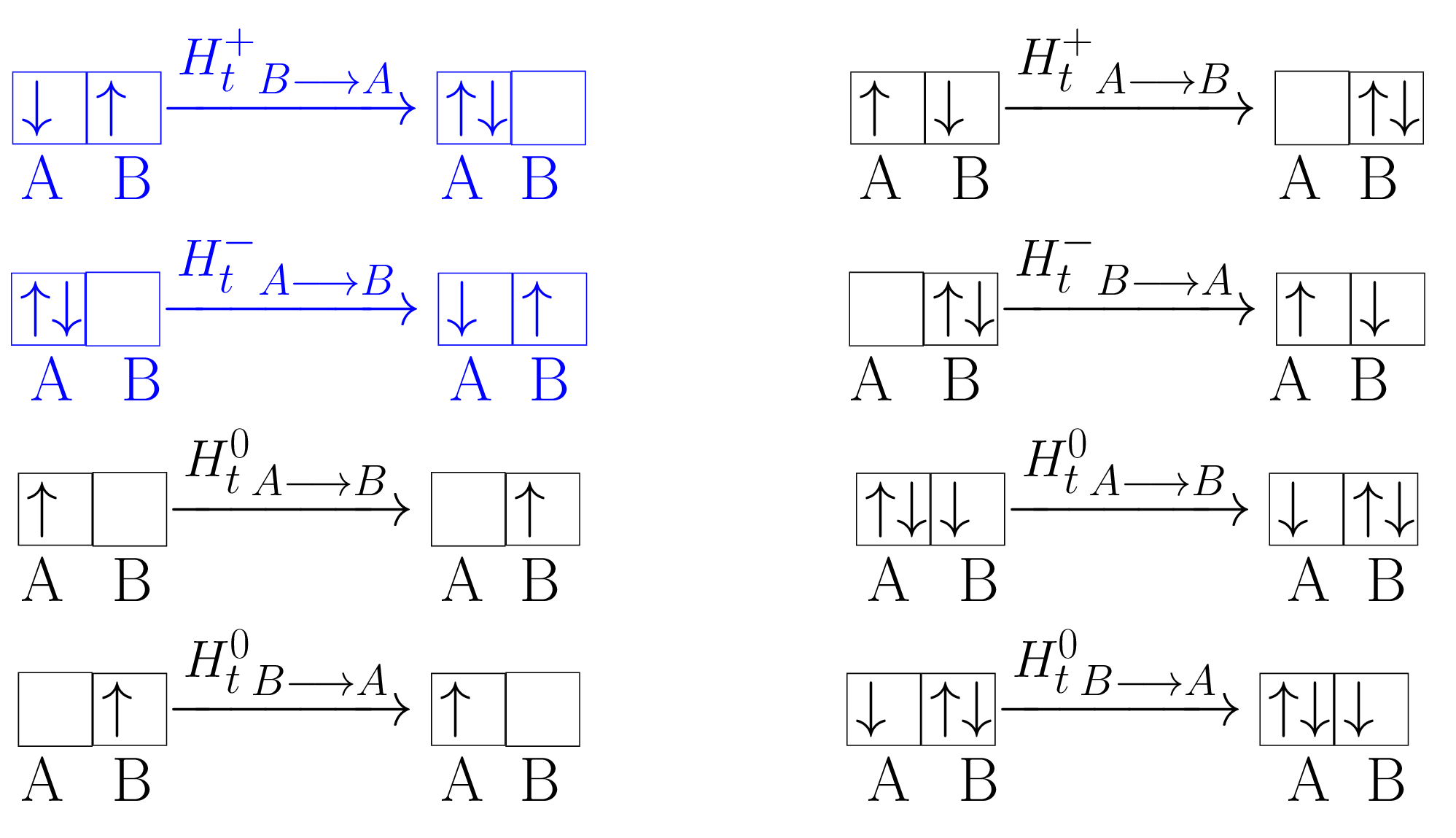}
 \hspace{2cm}   \caption{Nearest neighbour hopping processes for IHM.} 
\label{hopp_IHM} 
  \end{figure}

In the presence of non-zero hopping term, following nearest neighbour processes can take place as shown in Fig.~\ref{hopp_IHM}. 

\vspace{0.2cm}

$H_{t}^{+}$ processes involve the increase in double occupancy and hole occupancy by one, $H_{t}^{-}$ processes involve decrease in the double occupancy and hole occupancy by one and $H_{t}^{0}$ processes involve no change in the double occupancy or hole occupancy. Note that $H_{t B\rightarrow A}^{+}$ and $H_{t A\rightarrow B}^{-}$ are the only processes which are confined to the low energy sector of the Hilbert space. All other hopping processes mix the high energy and the low energy  part of the Hilbert space. 
Effective low energy Hamiltonian in the limit $U\sim \Delta \gg t$ can be obtained by doing similarity transformation which eliminates processes which interconnects the high and low energy sector of the Hilbert space. The effective Hamiltonian is give by
\be
 \mathcal{H}_{eff}=e^{iS}He^{-iS}= H+i[S,H]+\dfrac{i^2}{2}[S,[S,H]]+...   
\ee
Here, $S$, the transformation operator is perturbative in $t/\Delta$ and $t/(U+\Delta)$ and is given by
\be
    iS=\frac{1}{U+\Delta}({H_{t}^{+}}_{A\rightarrow B}-{H_{t}^{-}}_{B\rightarrow A })+\frac{1}{\Delta}({H_{t}^{0}}_{A\rightarrow B}-{H_{t}^{0}}_{B\rightarrow A})
 \ee
Higher order ($O(t^2/U)$) terms that arise from $[S,H_{t}]$ and $[S,[S,H_{0}]]$ and connects the low energy sector to the high energy sector can be eliminated by including a second similarity transformation $S^{'}$ such that $[S^{'},H_{0}]$ cancels those terms.
 The effective Hamiltonian which  does not involve mixing between low and high energy subspaces upto order $t^{2}$ is,
\bea
\mathcal{H}_{eff}=H_{0}+H_{1,low}+\frac{1}{U+\Delta}[{H_{t}^{+}}_{A\rightarrow B},{H_{t}^{-}}_{B\rightarrow A}] \nonumber \\
\hspace{-0.3cm}
+\frac{1}{\Delta}[{H_{t}^{0}}_{A\rightarrow B},{H_{t}^{0}}_{B\rightarrow A}]+O(t^3/U^2) ...
\label{Heff}
\eea

Here $H_0=U\sum_i n_{i\ua}n_{i\da}-\frac{\Delta}{2}\sum_{i\in A} n_{i}+\frac{\Delta}{2}\sum_{i \in B}n_i$ and $H_{1,low} = H_{t B\rightarrow A}^{+} + H_{t A\rightarrow B}^{-}$ is the hopping process in the low energy sector.
If we now confine to the low energy subspace, $\frac{1}{U+\Delta}[{H_{t}^{+}}_{A\rightarrow B},{H_{t}^{-}}_{B\rightarrow A}] \sim -\frac{1}{U+\Delta}{H_{t}^{-}}_{B\rightarrow A}{H_{t}^{+}}_{A\rightarrow B}$ because the first term in the commutator demands a doublon at site B and a hole at site $A$ which is energetically not favourable.
Similarly, $\frac{1}{\Delta}[{H_{t}^{0}}_{A\rightarrow B},{H_{t}^{0}}_{B\rightarrow A}]
\sim -\frac{1}{\Delta}{H_{t}^{0}}_{B\rightarrow A}{H_{t}^{0}}_{A\rightarrow B}$ because the first term in the commutator either demands a doublon at B or a hole at A and thus is not allowed because they belong to the high energy sector.

\subsection{Low energy Hamiltonian in terms of projected Fermions}

    Since holes on A sublattice and doublons on B sublattice belong to the high energy sector, we have projected them out from the low energy Hilbert space and introduced new projected operators,  
    \begin{equation}
        \tilde{c}_{A\sigma}^{\dagger}=\eta (\sigma)X_{A}^{d \leftarrow \bar{\sigma}}=c_{A\sigma}^{\dagger}n_{A\bar{\sigma}}
\label{cnewA}
    \end{equation}
     \begin{equation}
          \tilde{\tilde{c}}_{B\sigma}^{\dagger}=X_{B}^{\sigma \leftarrow 0}=c_{B\sigma}^{\dagger}(1-n_{B\bar{\sigma}})
\label{cnewB}
     \end{equation}
Note that $\{\tilde{\tilde{c}}_\sigma, \tilde{\tilde{c}}^\dagger_\sigma\} = 1-n_{\bar{\sigma}}$. 

While writing in terms of normal fermionic operators in the projected space, the order of the terms in the projected basis becomes important for A and B sublattices. On A sublattice, $\tilde{c}_{A\sigma}\tilde{c}_{A\sigma}^{\dagger}=\mathcal{P}_{h}c_{A\sigma}c_{A\sigma}^{\dagger}\mathcal{P}_{h}$ where as $\tilde{c}_{A\sigma}^{\dagger}\tilde{c}_{A\sigma} \neq\mathcal{P}_{h}c_{A\sigma}^{\dagger}c_{A\sigma}\mathcal{P}_{h}$. In the former case, both forms of operators count $\bar{\sigma}$ type single occupancies where as in the later case $\tilde{c}_{A\sigma}^{\dagger}\tilde{c}_{A\sigma}$ count  double occupancies while $c_{A\sigma}^{\dagger}c_{A\sigma}$ counts both double occupancies as well as $\sigma$ type single occupancies in the hole projected space.
 On B sublattice, the situation is opposite. $\tilde{c}_{B\sigma}^{\dagger}\tilde{c}_{B\sigma} =\mathcal{P}_{d}c_{B\sigma}^{\dagger}c_{B\sigma}\mathcal{P}_{d}$ and $\tilde{c}_{B\sigma}\tilde{c}_{B\sigma}^{\dagger}\neq\mathcal{P}_{d}c_{B\sigma}c_{B\sigma}^{\dagger}\mathcal{P}_{d}$. In the former case, both projected and normal fermionic operators count $\sigma$ type single occupancies where as in the later case the projected space operators count holes while normal fermionic representation count holes as well as $\bar{\sigma}$ type single occupancies in the doublon projected space.

\begin{figure}[h]
    \centering
    \includegraphics[width=7cm]{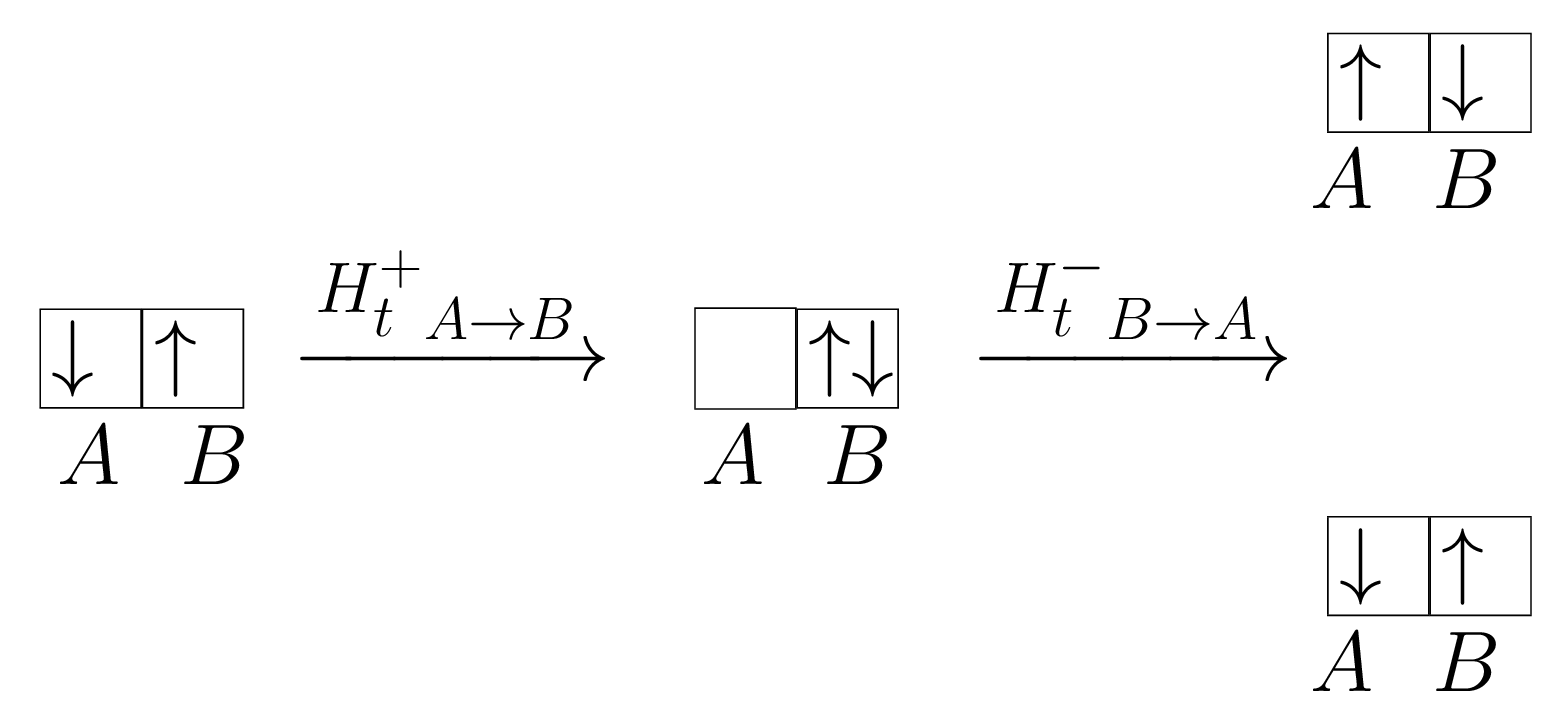}
    \caption{Spin exchange and spin preservation dimer terms for IHM.}
\label{spin_exch1}
\end{figure}

    In terms of new projected operators, $H_0$ in Eq.(~\ref{Heff}) can be written as $U\sum_{i \in A}(n_i-1)-\frac{\Delta}{2}[\sum_{i \in A}n_i-\sum_{i \in B}n_i]$. Here we have used that on a site $i \in A$, $n_{i\ua}n_{i\da} = n_i-1$ (see Eq.(~\ref{X_op})).  
 Since doublons have been projected out from B sublattice, in the low energy effective Hamiltonian there is no Hubbard term for B sublattice. The hopping term $H_{1,low}$ in the projected space does not involve holes on sublattice A and doublons on sublattice B. The representation in terms of projected operators is,

\bea
    H_{1,low}=-t\sum_{<ij>,\sigma}\tilde{c}_{iA\sigma}^{\dagger}\tilde{\tilde{c}}_{jB\sigma}+\tilde{\tilde{c}}_{jB\sigma}^{\dagger}\tilde{c}_{iA\sigma} \nonumber \\
= -t\sum_{<ij>,\sigma} \mathcal{P}[c^\dagger_{iA\sigma}c_{jB\sigma}+h.c.]\mathcal{P}
\label{ordert}
\eea
Here projection operator $\mathcal{P}$ projects out holes from the Hilbert space corresponding to sublattice A and doublons from the Hilbert space on sublattice B.   
\\
{\underline{\bf{$O(t^2/(U+\Delta))$ Dimer Terms}}}:
    In terms of Hubbard operators, the dimer term corresponding to
$\frac{1}{U+\Delta}[{H_{t}^{+}}_{A\rightarrow B},{H_{t}^{-}}_{B\rightarrow A}]\sim -\frac{1}{U+\Delta}{H_{t}^{-}}_{B\rightarrow A}{H_{t}^{+}}_{A\rightarrow B}$ becomes,
\[
H_{dimer}^1=  -\frac{t^2}{U+\Delta}\sum_{i \in A, j \in B,\sigma} [X_{i}^{\sigma \leftarrow \sigma} X_{j}^{\bar{\sigma}\leftarrow\bar{\sigma}}-X_{i}^{\bar{\sigma} \leftarrow \sigma} X_{j}^{\sigma\leftarrow\bar{\sigma}}] \] 
The corresponding process is represented in Fig.~[\ref{spin_exch1}]. In terms of projected fermionic operators, these dimer terms take the following form:
\[=-\dfrac{t^2}{U+\Delta}\sum_{i, j,\sigma}[\tilde{c}_{iA\bar{\sigma}}\tilde{c}_{iA\bar{\sigma}}^{\dagger}\tilde{\tilde{c}}_{jB\bar{\sigma}}^{\dagger}\tilde{\tilde{c}}_{jB\bar{\sigma}}-\tilde{c}_{iA\sigma}\tilde{c}_{iA\bar{\sigma}}^{\dagger}\tilde{\tilde{c}}_{jB\sigma}^{\dagger}\tilde{\tilde{c}}_{jB\bar{\sigma}}] \]
\be
= J_{1}\sum_{i, j}\mathcal{P}(S_{iA}.S_{jB}-(2-n_{iA})n_{jB}/4)\mathcal{P}
\ee

with $J_{1} = \f{2t^2}{U+\Delta}$. Projection operator $\mathcal{P}$ projects out hole from sublattice A and doublons from sublattice B. 
Note that in writing above renormalised form of the Heisenberg part of the Hamiltonian, we have imposed SU(2) symmetry by hand~\cite{GA,garg_nature}. Within simplest approximation of spin resolved densities being same in projected and unprojected states, the Gutzwiller approximation factor for $S_{iA}^zS_{jB}^z$ remains unity while the Gutzwiller factor for $S_{iA}^{+}S_{jB}^{-}+h.c.$ term is $g_s$. Since the original Hamiltonian is $SU(2)$ symmetric, the renormalised Hamiltonian obtained after taking into account the effect of projection, must also be $SU(2)$ symmetric. Hence we used $g_s$ to be the Gutzwiller factor for $S_{iA}^zS_{jB}^z$ term as well.     

\begin{figure}[h]
    \centering
    \includegraphics[width=7cm]{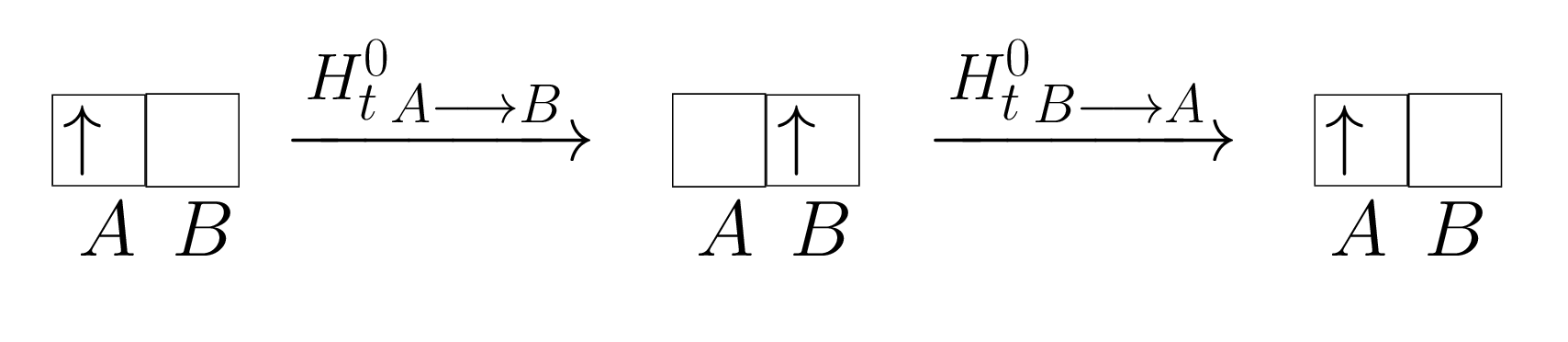}
    \includegraphics[width=7cm]{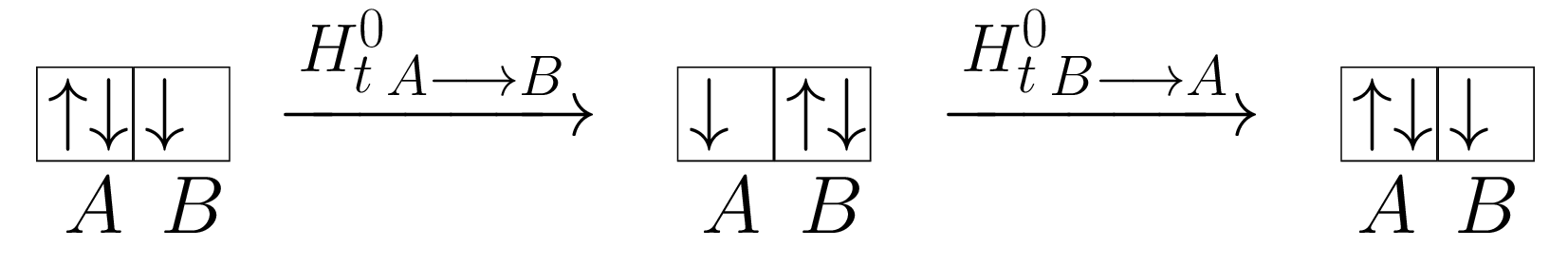}
    \caption{Top: Hopping of a single spin to site $B$ and back to site $A$. Bottom panel shows hopping of a doublon from $A$ to $B$ and back to $A$.}
    \label{dimer2}
\end{figure}

The dimer term corresponding to   $[{H_{t}^{0}}_{A\rightarrow B},{H_{t}^{0}}_{B\rightarrow A}]$  involves hopping of an e or a doublon from some site to its nearest neighbour site and back to the initial site as shown in Fig.~[\ref{dimer2}].

This process is of order $t^2/\Delta$ and can be written as 
\[
H_{dimer}^2=-\frac{t^2}{\Delta}\sum_{\sigma,<ij>}\big[X_{iA}^{\sigma\leftarrow \sigma}X_{jB}^{0\leftarrow0}+X_{iA}^{d\leftarrow d}X_{jB}^{\bar{\sigma}\leftarrow\bar{\sigma}}\big]\]
In terms of projected operators we get
\[=-\frac{t^2}{\Delta}\sum_{\sigma,<ij>}\big[\tilde{c}_{iA\bar{\sigma}}\tilde{c}_{iA\bar{\sigma}}^{\dagger}\tilde{\tilde{c}}_{jB\sigma}\tilde{\tilde{c}}_{jB\sigma}^{\dagger}+\tilde{c}_{iA\sigma}^{\dagger}\tilde{c}_{iA\sigma}\tilde{\tilde{c}}^\dagger_{jB\bar{\sigma}}\tilde{\tilde{c}}_{jB\bar{\sigma}}\big]\]
\be
=-\frac{t^2}{\Delta}\sum_{<ij>,\sigma}\mathcal{P}\big[(1-n_{iA\bar{\sigma}})(1-n_{jB})+(n_{iA}-1)n_{jB\bar{\sigma}}\big]\mathcal{P}
\label{hd_hopp}
\ee

\begin{figure}[h]
    \centering
    \includegraphics[width=8cm]{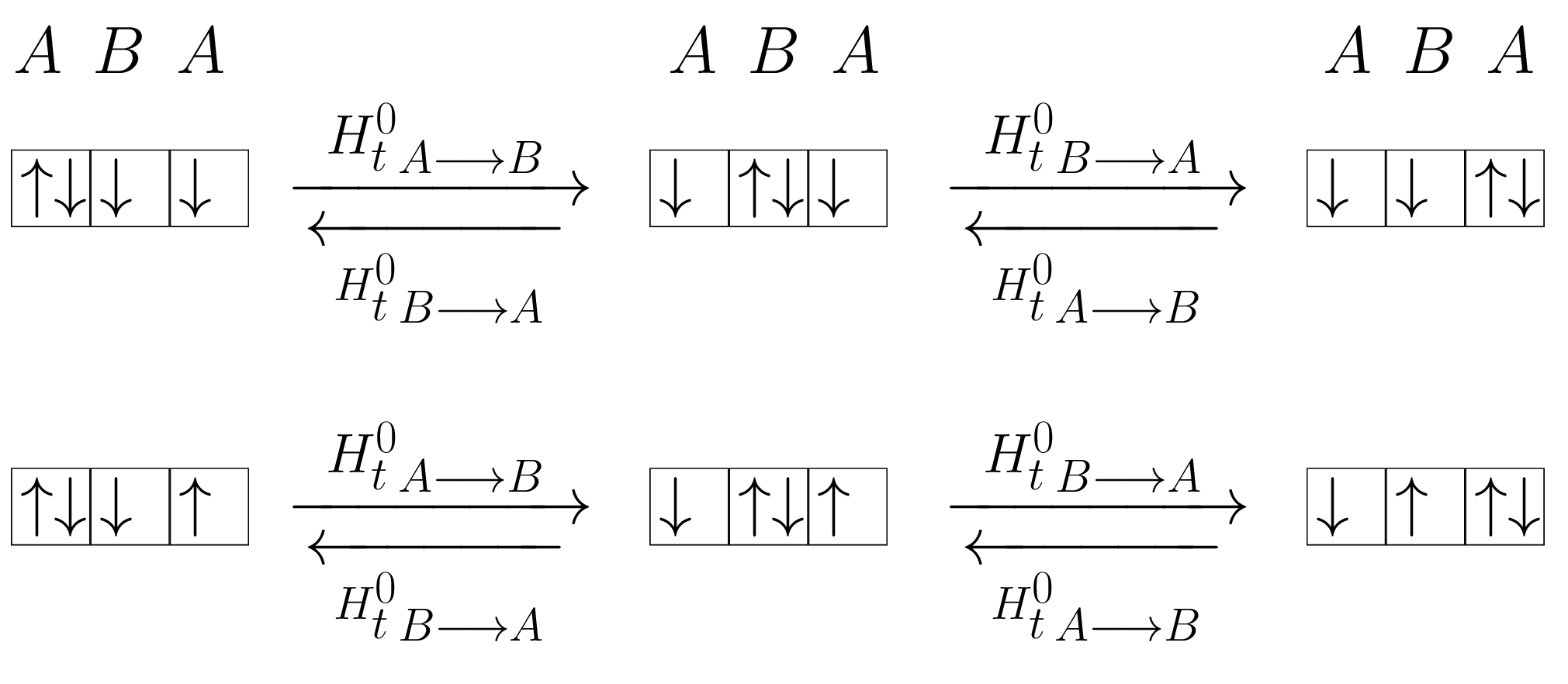}
    \caption{Effective next nearest neighbour hopping of a doublon within A sublattice.}
     \label{trimer1}
    \end{figure}

\begin{figure}[h]
    \centering
    \includegraphics[width=8cm]{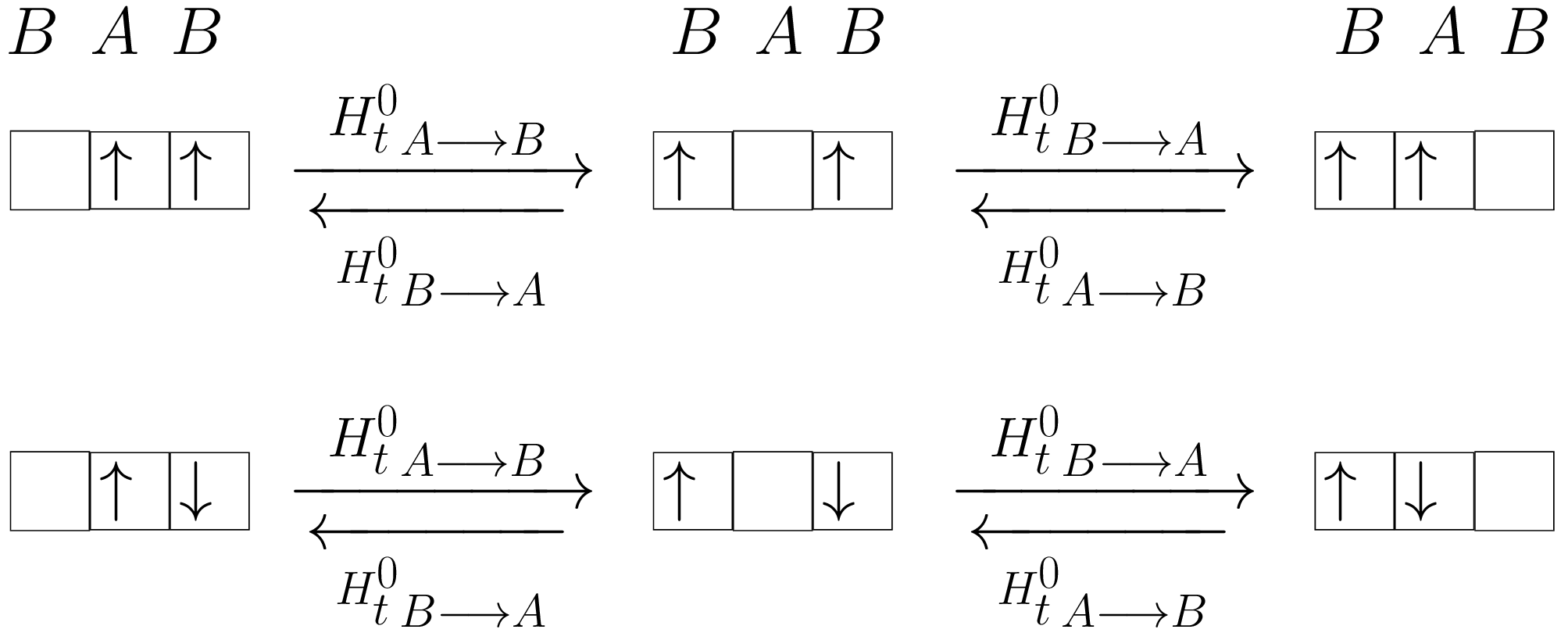}
    \caption{Effective next nearest neighbour hopping of hole for IHM.}
	\label{trimer2} 
   \end{figure}

\underline{\bf{$O(t^2/U)$ Trimer terms}}:

    Trimer terms involve hopping of a doublon or a hole from a site to it's next nearest neighbour site. Effectively there is doublon hopping which is intra A sublattice hopping denoted by $H_{hopp}^{AA}$ where as the hole hopping is intra B sublattice hopping ($H_{hopp}^{BB}$) as shown in Fig.~[\ref{trimer1},~\ref{trimer2}]. 

In terms of $X$ operators, hopping processes for doublon hopping, which is of $O(t^2/\Delta)$, on A sublattice are represented as, $H_{hopp}^{AA}=$
\[
-\frac{t^2}{\Delta}\sum_{\sigma,<ijk>}X_{kA}^{d\leftarrow\bar{\sigma}}X_{jB}^{\bar{\sigma}\leftarrow\bar{\sigma}}X_{iA}^{\bar{\sigma}\leftarrow d}+X_{kA}^{d\leftarrow \sigma}X_{jB}^{\sigma\leftarrow\bar{\sigma}}X_{iA}^{\bar{\sigma}\leftarrow d}+h.c.
\]
In terms of projected operators, it is represented as
  
\[
=-\frac{t^2}{\Delta}\sum_{\sigma,<ijk>}(\tilde{c}_{kA\sigma}^{\dagger}\tilde{\tilde{c}}_{jB\bar{\sigma}}^{\dagger}\tilde{\tilde{c}}_{jB\bar{\sigma}}\tilde{c}_{iA\sigma}+\tilde{c}_{iA\bar{\sigma}}\tilde{\tilde{c}}_{jB\bar{\sigma}}^{\dagger}\tilde{\tilde{c}}_{jB\sigma}\tilde{c}_{kA\sigma}^{\dagger})\]
\be
   =-\frac{t^2}{\Delta}\sum_{\sigma,<ijk>}\mathcal{P}(c^{\dagger}_{kA\sigma}n_{jB\bar{\sigma}}c_{iA\sigma}+
    c_{iA\bar{\sigma}}c^{\dagger}_{jB\bar{\sigma}}c_{jB\sigma}c^{\dagger}_{kA\sigma})\mathcal{P}
\label{intraA_hop}
    \ee
    
Similarly the hopping of holes within B sublattice, shown in Fig~[\ref{trimer2}], can be written in terms of $X$ operators as $H_{hopp}^{BB}=$
  \[
 -\frac{t^2}{\Delta}\sum_{\sigma,<jil>}X_{lB}^{0\leftarrow\sigma}X_{iA}^{\sigma\leftarrow\sigma}X_{jB}^{\sigma\leftarrow0}+X_{lB}^{0\leftarrow\bar{\sigma}}X_{iA}^{\bar{\sigma}\leftarrow\sigma}X_{jB}^{\sigma\leftarrow0}+ h.c.\]
which can be written in terms of projected operators as
\[
 =-\frac{t^2}{\Delta}\sum_{\sigma,<jil>}(\tilde{\tilde{c}}_{lB\sigma}\tilde{c}_{iA\bar{\sigma}}\tilde{c}_{iA\bar{\sigma}}^{\dagger}\tilde{\tilde{c}}_{jB\sigma}^{\dagger}+\tilde{\tilde{c}}_{jB\sigma}^{\dagger}\tilde{c}_{iA\sigma}\tilde{c}_{iA\bar{\sigma}}^{\dagger}\tilde{\tilde{c}}_{lB\bar{\sigma}})\]
\be
 =-\frac{t^2}{\Delta}\sum_{\sigma,<jil>}\mathcal{P}(c_{lB\sigma}[(1-n_{iA\bar{\sigma}})c_{jB\sigma}^{\dagger}+
     c_{iA\sigma}^{\dagger}c_{iA\bar{\sigma}}c_{jB\bar{\sigma}}^{\dagger}])\mathcal{P}
\label{intraB_hop}
\ee

\subsection{Gutzwiller approximation}
The effective low energy Hamiltonian obtained in the above section can be written as $\mathcal{H}_{eff} = \mathcal{P}\tilde{H}\mathcal{P}$ where $\mathcal{P}$ will project out holes from A sublattice and doublons from B sublattice for half-filling and densities close to half-filling. Within Gutzwiller approximation, the effect of this projection is taken approximately by renormalizing various coupling terms in $\tilde{H}$ by corresponding Gutzwiller factors such that eventually the expectation value of the renormalised Hamiltonian can be calculated in normal basis. 
Further we will calculate the Gutzwiller approximation factors under the assumption that the spin resolved densities before and after the projection remains same which will make Gutzwiller factors equal to $1$ for some terms in $\tilde{H}$. 
The renormalised Hamiltonian can be written as 
\[
\tilde{H}=H_0-t\sum_{\sigma,<ij>}g_{t\sigma}[{c}_{iA\sigma}^{\dagger}{c}_{jB\sigma} + h.c.] \]
\vskip-0.3cm
\[
-\frac{t^2}{\Delta}\sum_{<ij>,\sigma}[g_1(1-n_{iA\bar{\sigma}})(1-n_{jB})+g_{2}(n_{iA}-1)n_{jB\bar{\sigma}}] \]
\vskip-0.3cm
\[
-\frac{t^2}{\Delta}\sum_{\sigma,<ijk>}(g_{3\sigma}c^{\dagger}_{kA\sigma}n_{jB\bar{\sigma}}c_{iA\sigma}+g_4c_{iA\bar{\sigma}}c^{\dagger}_{jB\bar{\sigma}}c_{jB\sigma}c^{\dagger}_{kA\sigma} )+ h.c. \]
\vskip-0.3cm
\[
-\frac{t^2}{\Delta}\sum_{\sigma,<jil>}(g_{5\sigma} c_{lB\sigma}(1-n_{iA\bar{\sigma}})c_{jB\sigma}^{\dagger}+g_6 c_{lB\sigma}c_{iA\sigma}^{\dagger}c_{iA\bar{\sigma}}c_{jB\bar{\sigma}}^{\dagger})+ h.c. \]
\vskip-0.3cm
\be
+\frac{2t^2}{U+\Delta}\sum_{<i,j>}(g_{s}S_{iA}.S_{jB}-\f{1}{4}(2-n_{iA})n_{jB}) 
 \label{H_ren}
\ee
Here $g_{t,\sigma}$ and $g_s$ are Gutzwiller approximation factors for the nearest neighbour hopping and spin exchange terms. $g_1$ and $g_2$ are Gutzwiller factors for dimer terms $H_{dimer}^{1,2}$ respectively.  
\begin{figure*}[ht]
\hfill
\hspace{-3cm}
\subfigure{\frame{\includegraphics[width=6cm]{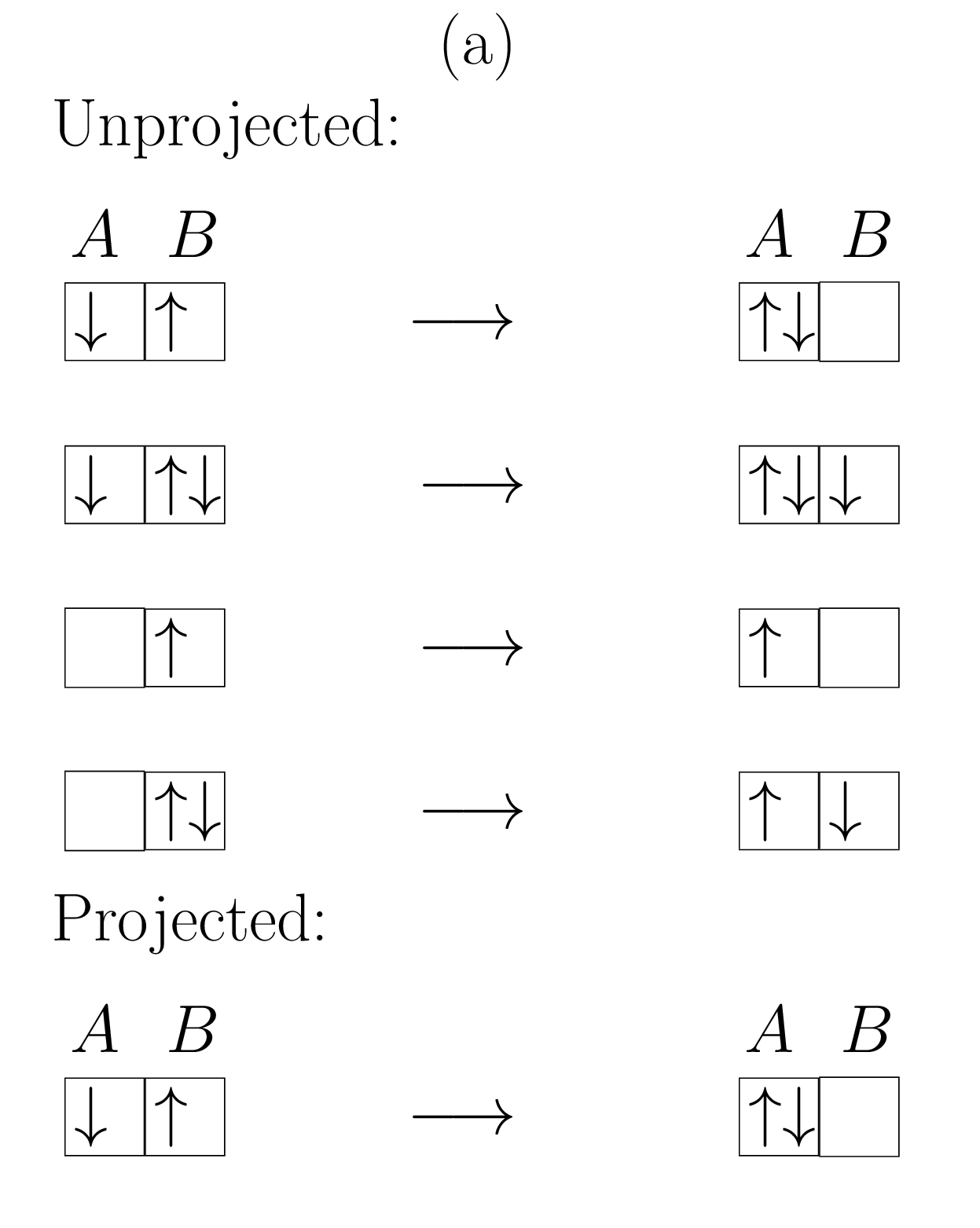}}}
\hfill
\subfigure{\frame{\includegraphics[width=6cm]{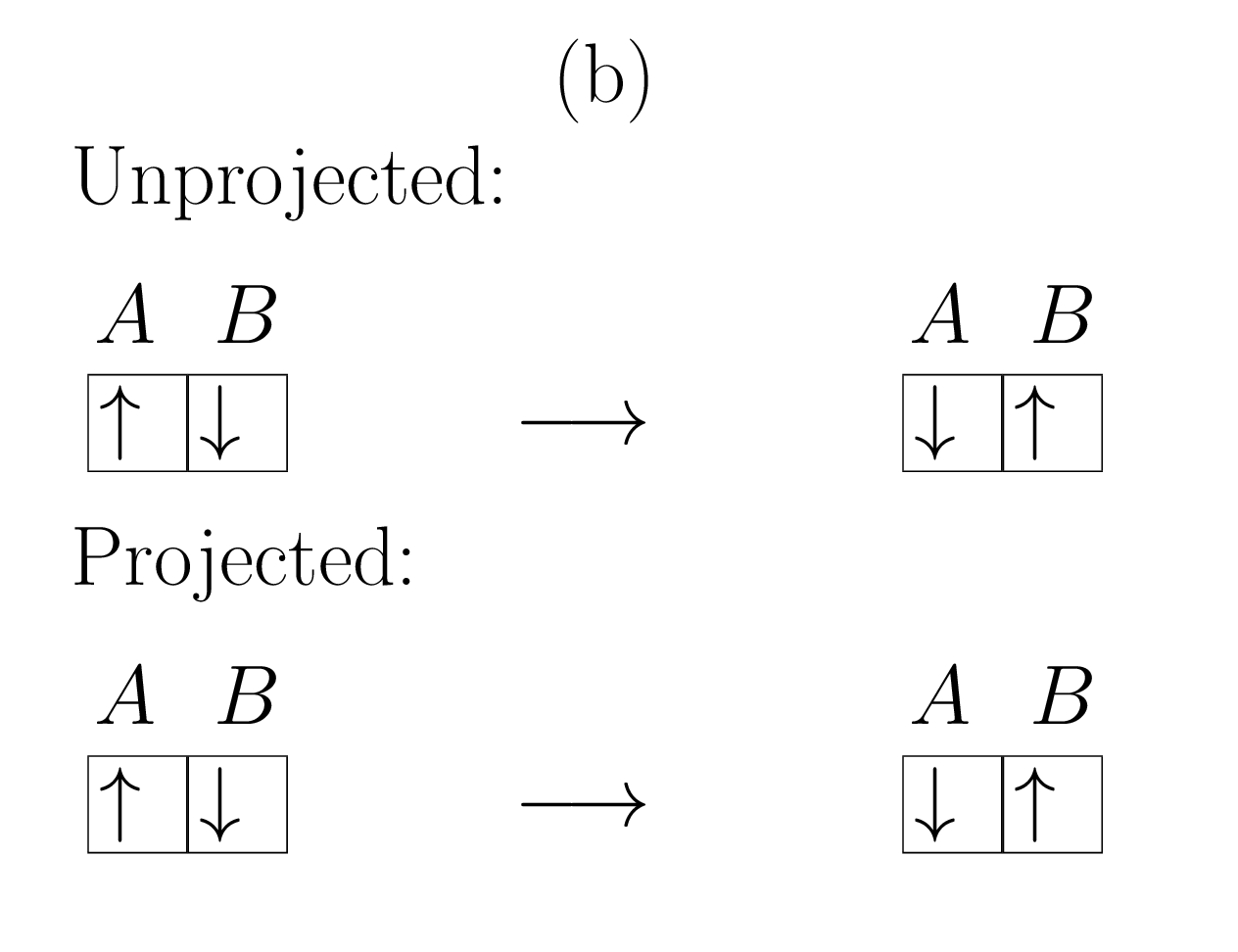}}}
\hfill
\caption{(a) Processes involved in the calculation of nearest neighbour hopping renormalization factor, $g_{t,\sigma}$. (b) Processes involved in the calculation of spin exchange renormalization factor $g_{s}$.}
\label{gt_gs}
\end{figure*}
$g_{3\sigma}$ and $g_4$ are Gutzwiller factors for intra sublattice hopping of doublons on A sublattice and $g_{5,\sigma}$ and $g_6$ are Gutzwiller factors for the intra sublattice hopping of holes on B sublattice. As we will demonstrate, some of the Gutzwiller factors are spin symmetric while other might be spin dependent in a spin symmetry broken phase like in anti-ferromagnetically ordered phase. Below we evaluate them one by one for various processes involved in $\mathcal{H}_{eff}$. 
We have enlisted below in Table [2] the probabilities of different states in the doublon projected basis.
Probabilities for various states for the hole projected sublattice were enlisted in Table.1.
{
\begin{center}
 \begin{tabular}{||c | c | c||}  
 \hline
 States & Unprojected  & Projected\\ [0.5ex] 
 \hline\hline
$ |\uparrow\rangle$ & ${{\bf {n}}_{\uparrow}(1 - {\bf{n}}_{\downarrow})}$ &  ${\bf{n}}_{\uparrow}$ \\ \hline
 $|\downarrow\rangle$ & ${{\bf{n}}_{\downarrow}(1 - {\bf{n}}_{\uparrow})}$ &  ${\bf{n}}_{\downarrow}$ \\ 
 \hline
$ |\uparrow\downarrow\rangle$  & ${\bf{n}}_{\uparrow}{\bf{n}}_{\downarrow}$ & $0$ \\ 
\hline
$|0\rangle$ & $(1 - {\bf{n}}_{\uparrow})(1 - {\bf{n}}_{\downarrow})$ & $(1-{\bf{n}})$ \\ [1ex] 
 \hline
\end{tabular}
\end{center}
}

\normalsize

Table 2. Probabilities of different states in terms of electron densities in unprojected and  doublon projected basis.
 \\
\vskip0.3cm
    \begin{figure*}[ht]
\hfill
\hspace{-3cm}
\subfigure{\frame{\includegraphics[width=6cm]{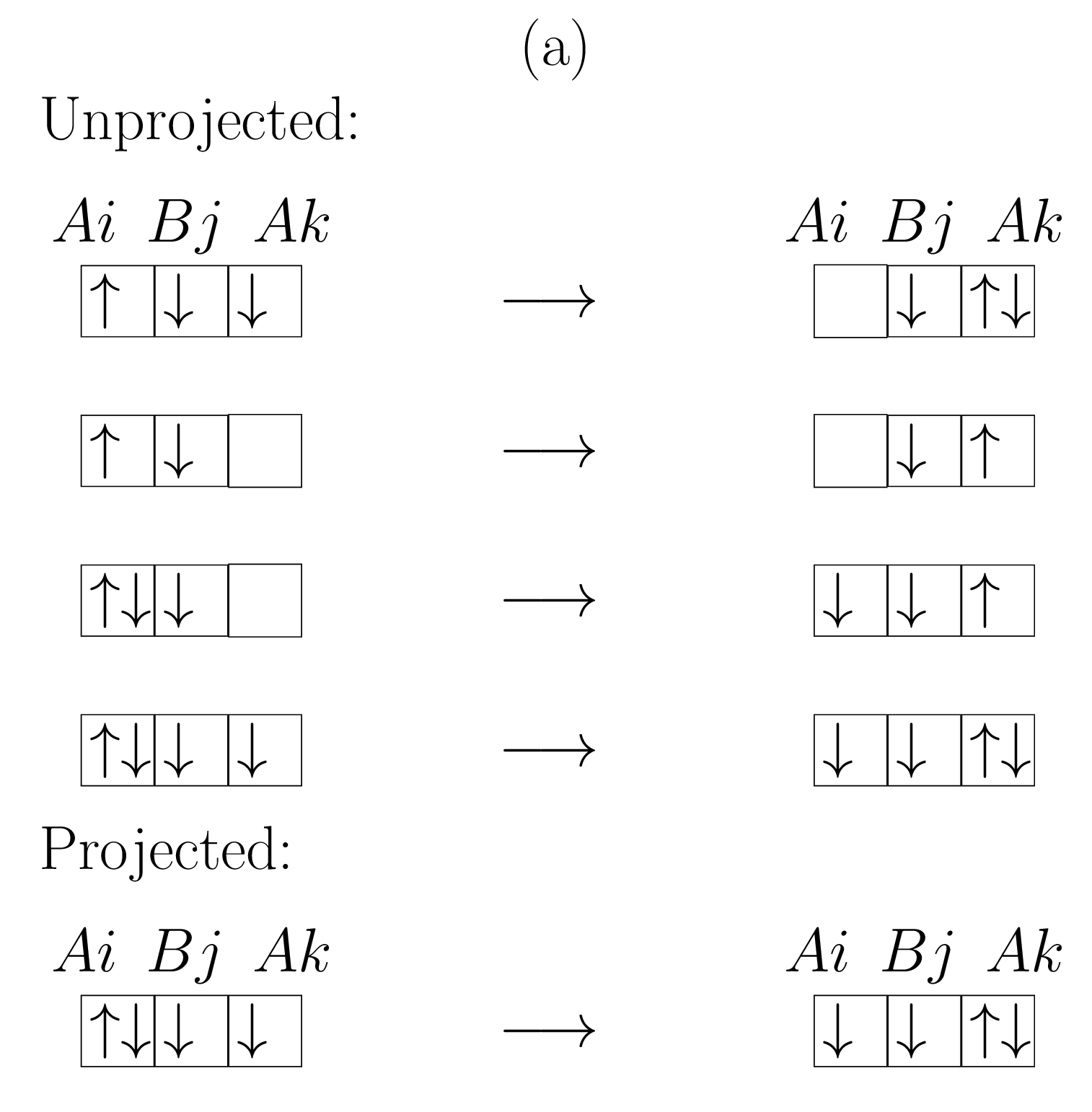}}}
\hfill
\subfigure{\frame{\includegraphics[width=6cm]{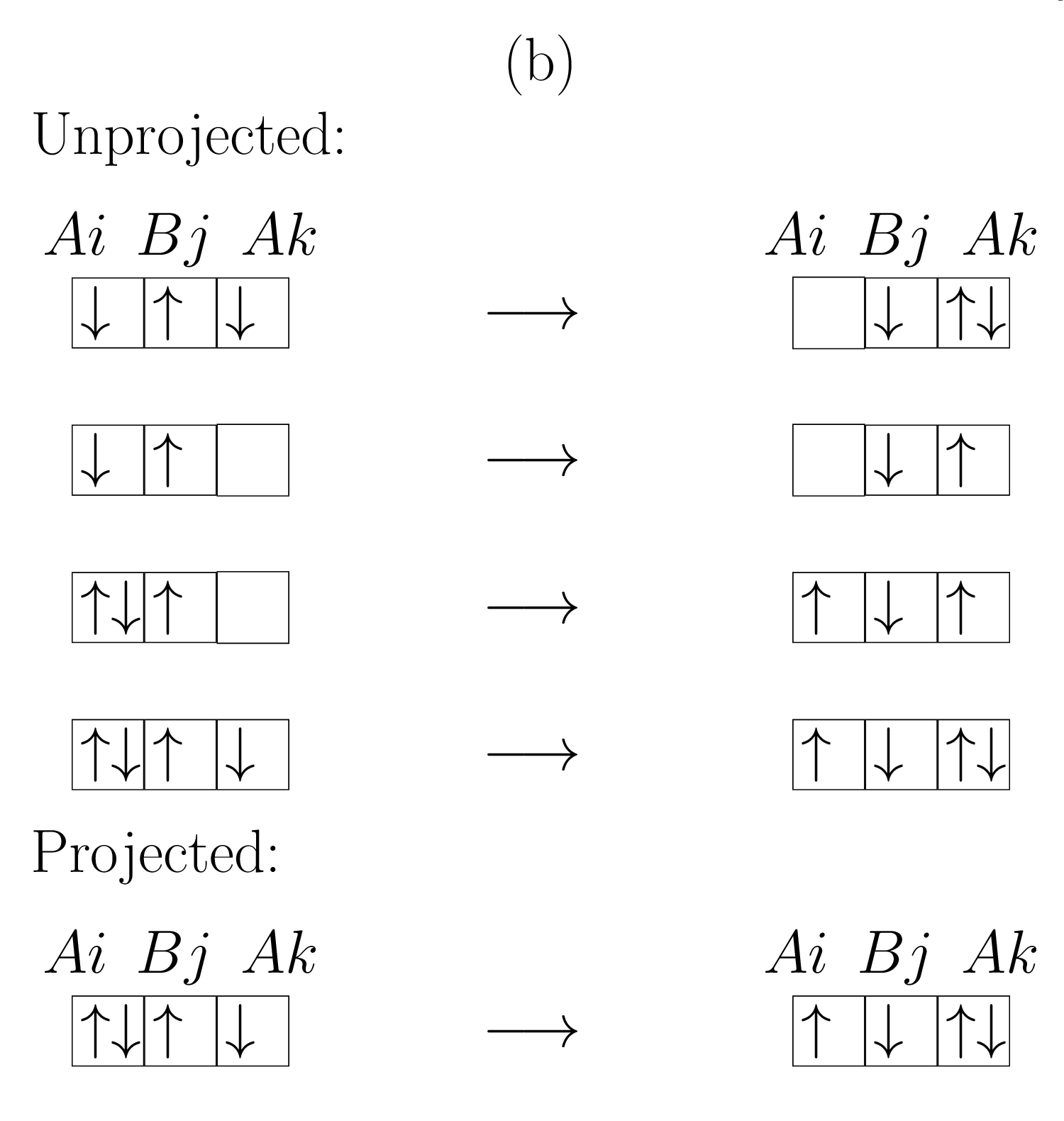}}}
\hfill
\caption{(a) Processes involved in the calculation of  $g_{3}$. Similar physical processes with doublon at  B site in the unprojected basis are considered in the calculation (but now shown here). (b) Processes involved in the calculation of $g_{4}$.}
\label{g3g4}
\end{figure*}

As we mentioned earlier, this analysis holds at half-filling and for densities not far from half-filling. Even if the system is overall half-filled, the individual sublattices are not, A  subalttice is electron doped where as B sublattice is hole doped. At half-filling in the Hubbard model, Gutzwiller renormalization factor for hopping is zero because the system is an Antiferromagnetic Mott Insulator where as in case of IHM, the density difference between the sublattices result in finite $g_{t,\sigma}$. Here, as we will show, the density difference between two sublattices plays the role of doping in case of Hubbard model. Also, the trimer terms are present in half-filled IHM which result in intra sublattice hopping of holes and doublons where as half-filled Hubbard model has no trimer terms. 

\begin{figure*}[ht]
\hfill
\hspace{-3cm}
\subfigure{\frame{\includegraphics[width=6cm]{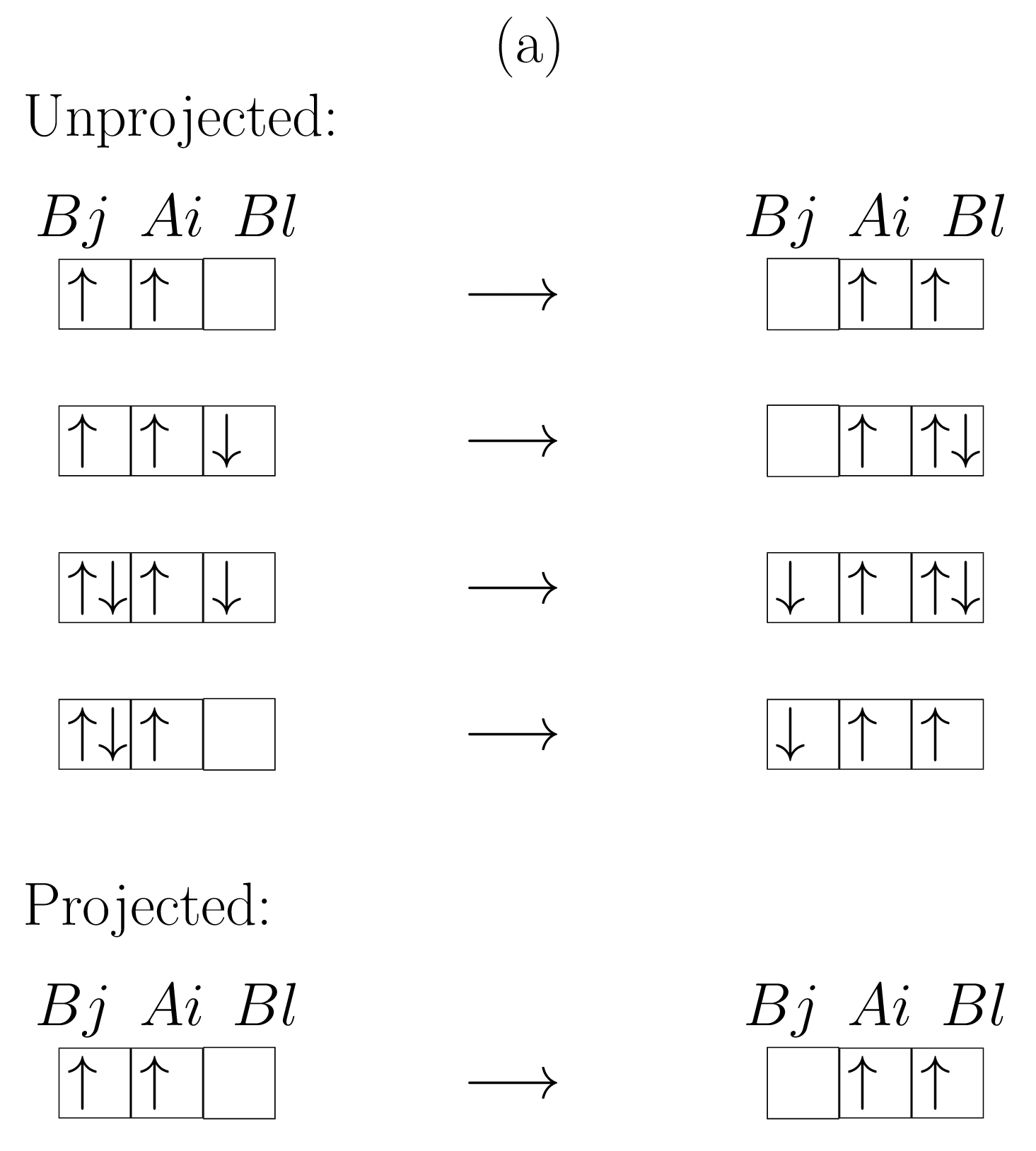}}}
\hfill
\subfigure{\frame{\includegraphics[width=6cm]{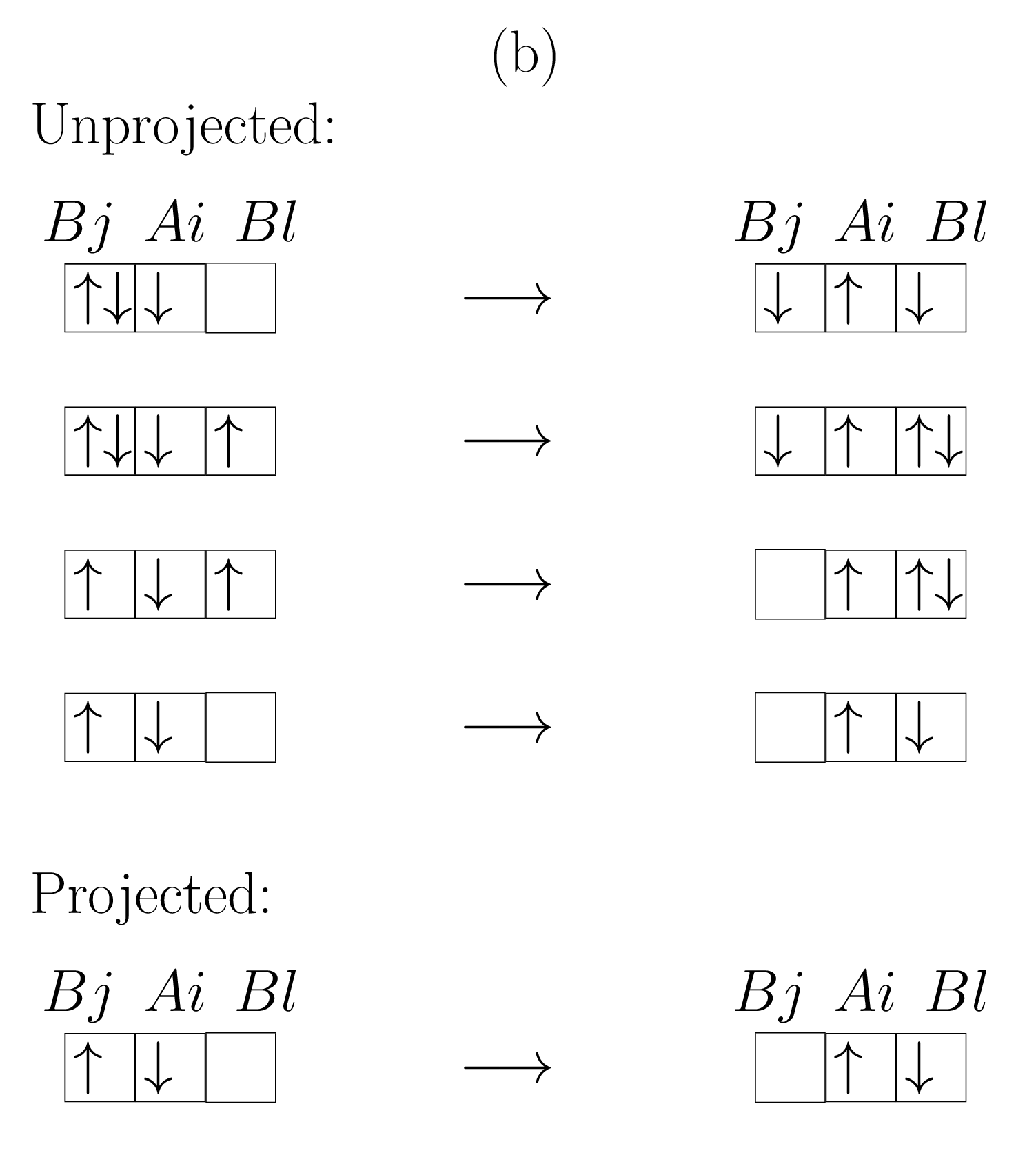}}}
\hfill
\caption{(a) Processes involved in the calculation of $g_{5}$. Similar  physical processes with hole at A site in the unprojected basis are considered in the calculation of $g_5$ (but not shown here). (b) Processes involved in the calculation of $g_{6}$.}
\label{g5g6}
\end{figure*}

Below we first give the general expression for $g_{t,\sigma}$ and $g_s$ at any filling and then evaluate them for special case of half filling, $\dfrac{{\bf{n}}_A+{\bf{n}}_B}{2}=1$.
The probability of nearest neighbour hopping of an $\ua$ electron in the unprojected space (shown in Fig.~[\ref{gt_gs}]) is $(1-{\bf{n}}_{A\uparrow}){\bf{n}}_{B\uparrow}{\bf{n}}_{A\uparrow}(1-{\bf{n}}_{B\uparrow})$ and in the unprojected space it is $(1-{\bf{n}}_{A\uparrow}){\bf{n}}_{B\uparrow}({\bf{n}}_{A}-1)(1-{\bf{n}}_{B})$. Then, the Gutzwiller renormalization factor, 
\be
g_{t\uparrow}=\sqrt{\dfrac{({\bf{n}}_A-1)(1-{\bf{n}}_B)}{{\bf{n}}_{A\uparrow}(1-{\bf{n}}_{B\uparrow})}}
\label{gt}
\ee
 Let $\delta=\dfrac{{\bf{n}}_A-{\bf{n}}_B}{2}$ be the density difference between two sublattices. Then at half-filling, density of A sublattice is ${\bf{n}}_A=1+\delta$ and that on B sublattice is $ {\bf{n}}_B=1-\delta$. Let the magnetization on A sublattice, $m_A=\bf{n}_{A\uparrow}-\bf{n}_{A\downarrow}$, then at half-filling due to particle-hole symmetry, $ m_A=-m_B=m$.
One can re-write $g_{t,\sigma}=\dfrac{2\delta}{1+\delta+\sigma m}$ in an anti-ferromagnetically ordered phase at half-filling. For $m=0$, $g_t$ takes the form similar to that known for doped $t-J$ model with $\delta$ , the density difference in IHM, playing the role of hole doping in $t-J$ model.

Now consider the spin exchange process shown in Fig.~[\ref{gt_gs}(b)]. The probability for this process to take place in the unprojected basis is ${\bf{n}}_{A\uparrow}(1-{\bf{n}}_{A\downarrow}){\bf{n}}_{B\downarrow}(1-{\bf{n}}_{B\uparrow}){\bf{n}}_{A\downarrow}(1-{\bf{n}}_{A\uparrow}){\bf{n}}_{B\uparrow}(1-{\bf{n}}_{B\downarrow})$ where as in the projected basis it is $(1-{\bf{n}}_{A\downarrow}){\bf{n}}_{B\downarrow}(1-{\bf{n}}_{A\uparrow}){\bf{n}}_{B\uparrow}$, resulting in the Gutzwiller factor,
\be
g_s=\sqrt{\dfrac{1}{{\bf{n}}_{A\uparrow}{\bf{n}}_{A\downarrow}(1-{\bf{n}}_{B\uparrow})(1-{\bf{n}}_{B\downarrow})}}
\label{gs}
\ee
Again at half-filling in an AFM ordered phase $g_s= 4/((1+\delta)^2-m^2)$ which for $m=0$ again maps to the $g_s$ factor for doped $t-J$ model with $\delta$ playing the role of hole-doping in that case.

Gutzwiller factors $g_{1}$ , $g_{2}$ are $1$ because dimer terms $H_{dimer}^{1,2}$ are product of densities. Under the assumption that the spin resolved unprojected and projected densities are same, the Gutzwiller factors for these terms are 1.

 Now we will calculate Gutzwiller factors for various trimer terms shown in Fig.[\ref{trimer1}] and Fig.~[\ref{trimer2}]. Fig.[\ref{g3g4}(a)] shows hopping of an $\ua$ electron within A sublattice with a spin ($\da$) on the intermediate B site being preserved. In the unprojected basis, the probability for this process to happen is ${\bf{n}}_{A\uparrow}^{2}(1-{\bf{n}}_{A\uparrow})^{2}{\bf{n}}_{B\downarrow}^{2}$. It is to be noted that processes with either a down type particle or a doublon at the intermediate B site have been considered in the unprojected space. Like wise, the probability for the process to happen in the projected basis is $({\bf{n}}_{A}-1)^{2}(1-{\bf{n}}_{A\uparrow})^{2}{\bf{n}}_{B\downarrow}^{2}$. Therefore, the Gutzwiller factor for this process is
\be
g_{3\uparrow}=\dfrac{{\bf{n}}_A-1}{{\bf{n}}_{A\uparrow}}=\dfrac{2\delta}{1+\delta+m}
\label{g3}
\ee
where the expression on right most side holds in case of half-filling for a non-zero staggered magnetisation. In general one gets $g_{3\sigma} = \frac{{{\bf{n}}}_A-1}{{\bf{n}}_{A\sigma}}$. 
Fig.~[\ref{g3g4}(b)] depicts hopping processes on A sublattice in which spin on the intermediate B site gets flipped. The probability in the unprojected basis for this process to occur is $(1-{\bf{n}}_{A\uparrow})(1-{\bf{n}}_{A\downarrow}){\bf{n}}_{A\uparrow}{\bf{n}}_{A\downarrow}(1-{\bf{n}}_{B\uparrow})(1-{\bf{n}}_{B\downarrow}){\bf{n}}_{B\uparrow}{\bf{n}}_{B\downarrow}$ where as that in the projected basis is $({\bf{n}}_{A}-1)^{2}(1-{\bf{n}}_{A\uparrow})(1-{\bf{n}}_{A\downarrow}){\bf{n}}_{B\uparrow}{\bf{n}}_{B\downarrow}$ resulting in the Gutzwiller factor 
\be
g_4=\dfrac{{\bf{n}}_A-1}{\sqrt{{\bf{n}}_{A\uparrow}{\bf{n}}_{A\downarrow}(1-{\bf{n}}_{B\uparrow})(1-{\bf{n}}_{B\downarrow})}}=\dfrac{4\delta}{(1+\delta)^2-m^2}
\label{g4}
\ee

Now consider the hopping processes within B sublattice depicted in Fig.~[\ref{trimer2}]. Fig.~[\ref{g5g6}[a]] shows hopping of an $\ua$ spin particle within B sublattice such that spin on the intermediate A site is preserved. Here, again it must be noted that processes with either an up particle or a hole at the intermediate A site has been considered in the unprojected basis.
In the unprojected basis the probability of this process is $(1-{\bf{n}}_{A\downarrow})^{2}{\bf{n}}_{B\uparrow}^{2}(1-{\bf{n}}_{B\uparrow})^{2}$ and that in the projected basis is $(1-{\bf{n}}_{A\downarrow})^{2}{\bf{n}}_{B\uparrow}^{2}(1-{\bf{n}}_{B})^{2}$  leading to the Gutzwiller factor 
\be
g_{5\uparrow}=\dfrac{1-{\bf{n}}_B}{1-{\bf{n}}_{B\uparrow}}=\dfrac{2\delta}{1+\delta+m}
\label{g5}
\ee
In general, $g_{5,\sigma} = (1-{\bf{n}}_B)/(1-{\bf{n}}_{B\sigma})$ is spin dependent. 

Another hopping process within B sublattice is the one in which spin on the intermediate A site gets flipped.  The probability for this process to occur in the unprojected basis is 
$(1-{\bf{n}}_{A\uparrow})(1-{\bf{n}}_{A\downarrow}){\bf{n}}_{A\uparrow}{\bf{n}}_{A\downarrow}(1-{\bf{n}}_{B\uparrow})(1-{\bf{n}}_{B\downarrow}){\bf{n}}_{B\uparrow}{\bf{n}}_{B\downarrow}$ and in projected space it is $(1-{\bf{n}}_{A\uparrow})(1-{\bf{n}}_{A\downarrow}){\bf{n}}_{B\uparrow}{\bf{n}}_{B\downarrow}(1-{\bf{n}}_{B})^{2}$. The Gutzwiller factor is therefore
\be
g_6=\dfrac{1-{\bf{n}}_B}{\sqrt{{\bf{n}}_{A\uparrow}{\bf{n}}_{A\downarrow}(1-{\bf{n}}_{B\uparrow})(1-{\bf{n}}_{B\downarrow})}}=\dfrac{4\delta}{(1+\delta)^2-m^2}
\ee

\subsection{Results for strongly correlated limit of IHM}
In this section we present results for the IHM in the limit $U\sim \Delta \gg t$ at half filling. To be specific, we do mean field decomposition of the renormalised low energy Hamiltonian in Eq.~(\ref{H_ren}) giving non zero expectation values to the following mean fields : (i)magnetization on A sublattice (B sublattice),$m_A$($m_B$) (ii)inter sublattice fock shift ($\chi_{AB}$) (iii)intra sublattice fock shift (iv) Hartree shifts and (v) the density difference between the two sublattices ($\delta$). The quadratic mean field Hamiltonian is solved by appropriate canonical transformation and mean fields are obtained self-consistently. Below we first provide a comparison of our approach with the results obtained from an exact diagonalisation (ED) study of this model for a one dimensional chain  followed up by the results towards a possible phase diagram of the IHM in the limit of validity of this approach.  
\subsubsection{{\bf{Comparison with ED results}}}
Below we first benchmark our approach of hole and doublon projection, implemented at the level of renormalised low energy Hamiltonian via Gutzwiller approximation, by comparing our results for 1-d chain with those obtained from exact diagonalization by Anusooya-Pati et. al~\cite{Soos}.
Since the formalism we have developed in this paper is valid for the regime of both $U$ and $\Delta$ being much larger than the hopping amplitude $t$ we compare our results for the largest value of $U$ for which results are shown in ~\cite{Soos}. 
\begin{figure}[h!]
\begin{center}
\includegraphics[width=6cm,angle=-90]{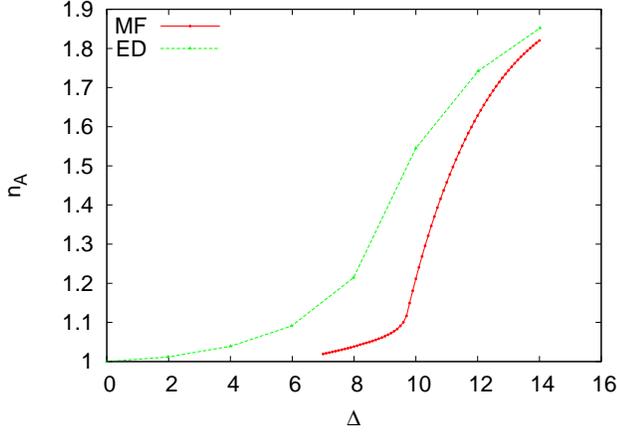}
\caption{Density on sublattice $A$ as a function of $\Delta$ for $U=10t$. ED results shown are obtained from ~\cite{Soos}.}
\label{nA_plot}
\end{center}
\end{figure}
Fig. ~\ref{nA_plot} shows the density on sublattice $A$ as a function of $\Delta$ for $U=10t$ for 1-d chain. ED result, obtained by digitizing the plot from work of Anusooya-Pati et. al~\cite{Soos}, is an extrapolation of finite size chains in the thermodynamic limit. For smaller values of $\Delta$ our formalism does not hold and hence the comparison has been shown for $\Delta\ge 7t$. The qualitative trend in both the calculations is same and as $\Delta$ increases better quantitative consistency is observed between the two calculations. Note that there is an overall factor of 2 difference in the ionic potential term in our Hamiltonian and the one used in Anusooya-Pati et.al. 
 After the checks to validate our formalism, we provide below the details of the phase diagram of IHM in the limit under consideration.
\subsubsection{{\bf{Phase diagram of IHM for $U\sim \Delta \gg t$}}}
The phase diagram of IHM in the limit $U\sim \Delta \gg t$ has not been explored in detail so far. Though there are a few numerical results available ~\cite{Soos,Soumen} but a complete understanding has been lacking mainly because no perturbative calculation has been developed in this limit so far. One of the reason is that the formalism for hole projection, which is essential in this limit, was missing so far in the literature. Below we provide details of various physical quantities based on the mean field analysis of our renormalised Hamiltonian for a 1-d chain and also discuss possible phases in higher dimensional cases. 
\begin{figure}[h!]
\begin{center}
\hspace{1.5cm}
\includegraphics[width=6.5cm,angle=-90]{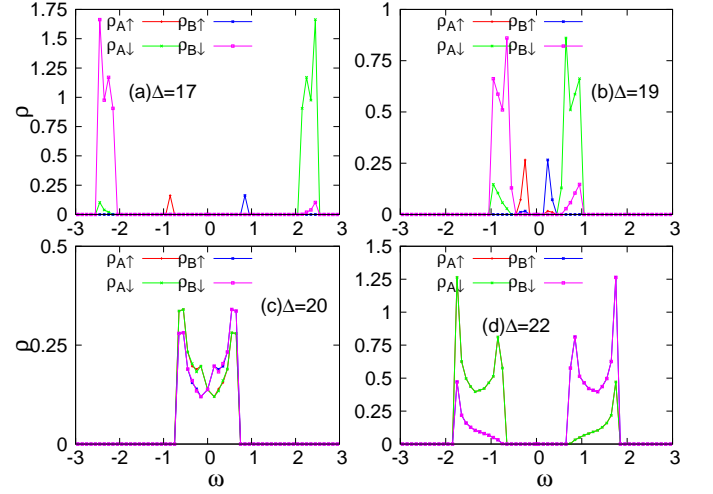}
\caption{Single particle DOS for $U=20t$ for a few values of $\Delta$. For $\Delta < U$, the system has spin asymmetry with $\rho_{\alpha\ua}(\omega) \ne \rho_{\alpha\da}(\omega)$. Also the gap in the DOS is larger for the down-spin channel. As $\Delta$ increases towards $U$, both the gaps decrease eventually giving a metallic phase for $\Delta\sim U$. As $\Delta$ increases further again the system becomes an insulator which has spin symmetry.}
\label{dos}
\end{center}
\end{figure}

{\bf{Single particle density of states}}: In this section we discuss the single particle density of states (DOS) $\rho_{\alpha\sigma}(\omega) \equiv \ -\sum_k Im~\hat{G}_{\alpha\sigma}(k,\omega^{+})/\pi$. Here $\alpha$ represents the sub lattice $A,B$ and $\sigma$ is the spin. In the Gutzwiller approximation, we must rescale the Green’s function $G_{\alpha\sigma}(k, \omega)$ with correct Gutzwiller factor~\cite{AG_NP} just like we did for hopping, spin exchange term and trimer terms in the Hamiltonian. Thus the renormalised $G_{\alpha\sigma}(k,\omega)= g_{t,\sigma}G^0_{\alpha\sigma}(k,\omega)$ where $G^0_{\alpha\sigma}(k,\omega)$ is the Green's functions calculated in the unprojected ground state of the Hamiltonian in Eq.~(\ref{H_ren}). The corresponding spectral function which is the imaginary part of the Green's function also satisfy the relation $A_{\alpha\sigma}(k,\omega)=g_{t\sigma}A^0_{\alpha\sigma}(k,\omega)$ resulting in the same relation for the single particle density of states $\rho_{\alpha\sigma}(\omega)$. 
\begin{figure}[h!]
\begin{center}
\includegraphics[width=6cm,angle=-90]{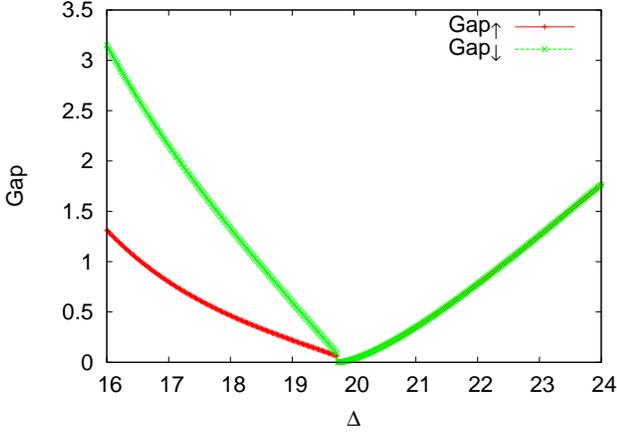}
\caption{The gap in the single particle density of states vs $\Delta$ for $U=20$. For $\Delta < U$, $gap_{\da} > gap_{\ua}$ and both decrease with incraese in $\Delta$ eventually becoming zero for $\Delta \sim U$. As $\Delta$ increases further, the gap opens up again but the gap in the up and down channel are equal in this phase.}
\label{gap}
\end{center}
\end{figure}
Fig.~\ref{dos} shows the renormalised single particle DOS in the projected Hilbert space for the IHM for $U=20t$ and a few values of $\Delta \sim U$. For $\Delta < \Delta_c$, the system has spin asymmetry as seen in the top two panels of Fig.~\ref{dos}. There is a gap in the DOS for both the up and the down spin channel, with gap in the up spin channel being smaller than that for the down spin channel. Both the gaps reduce with increase in $\Delta$ as shown in panels (b) and (c) of Fig.~\ref{dos} eventually becoming vanishingly small for a range of $\Delta$ values close to $\Delta=U$ where the system is metallic. On further increasing $\Delta$, the gap in the DOS opens up again but now the system is spin symmetric with both the gaps being equal. Fig.~\ref{gap} shows the behaviour of $gap_{\sigma}$ as a function of $\Delta$ for $U=20t$. 

\begin{figure}[h!]
\begin{center}
\includegraphics[width=6cm,angle=-90]{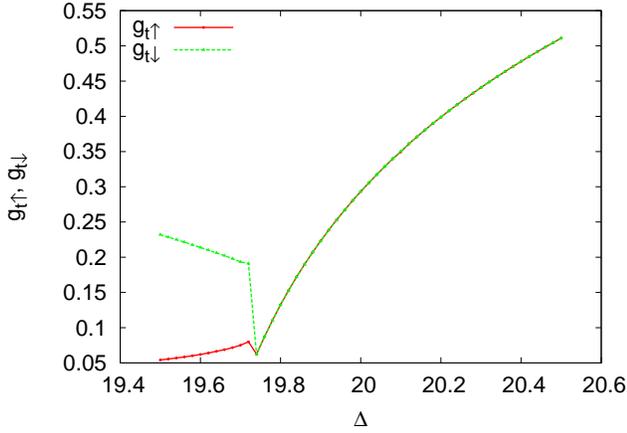}
\caption{Plot of $g_{t\sigma}$ vs $\Delta$ for $U=20$. In the metallic phase $g_{t\sigma}$ provides the quasi-particle weight.}
\label{Z}
\end{center}
\end{figure}

Existence of a metallic phase intervening the two insulating phases of the IHM has been a debatable issue in the literature. Though the solution of DMFT self consistent equations in the paramagnetic (PM) sector at half filling at zero temperature shows an intervening metallic phase~\cite{AG1}, in the spin asymmetric sector, the transition from paramagnetic band insulator (PM BI) to anti-ferromagnetic (AFM) insulator preempts the formation of a para-metallic phase~\cite{kampf,cdmft}. But determinantal quantum Monte-Carlo results demonstrated the presence of a metallic phase even in spin asymmetric solution~\cite{qmc_ihm1,qmc_ihm2}.  Exact diagonalization for 1d chains ~\cite{Soos} have also shown signatures of the presence of a metallic phase via calculation of the charge stiffness.
In all the cases, where an intervening metallic phase has been demonstrated, it was also shown that the width of the metallic phase shrinks with increase in $U$ and $\Delta$. A very narrow metallic regime observed in our approach for the IHM at half filling for $U \sim \Delta \gg t$ is completely consistent with these studies. 
 
The renormalised momentum distribution function  $n_{\alpha\sigma}(k) = \int d\omega A_{\alpha\sigma}(k,\omega) = g_{t\sigma}n^0_{\alpha\sigma}(k)$, where $n^0_{\alpha\sigma}(k)$ is the momentum distribution function in the unprojected Hilbert space. Thus the quasi-particle weight, which is the jump in the momentum distribution function at the Fermi momentum, is $Z=g_{t\sigma}$. Fig.~\ref{Z} shows $g_{t\sigma}$ vs $\Delta$ for $U=20t$. In the metallic regime, that is, for $\Delta \sim 20t$, $g_{t\ua}= g_{t\da} \ll 1$ which indicates that we actually have a {\it bad} metal, with very heavy quasi-particles, intervening the two insulators. Note that in the insulating regime $g_{t\sigma}$ does not carry the meaning of quasi-particle weight. 

\begin{figure}[h!]
\begin{center}
\includegraphics[width=6cm,angle=-90]{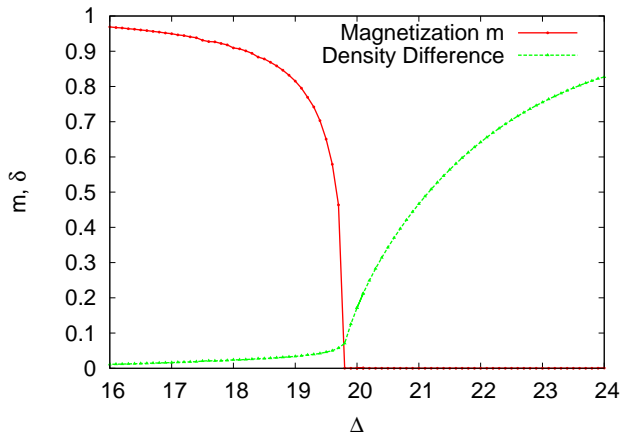}
\caption{ Staggered magnetisation $m$ and staggered density $\delta$ vs $\Delta$ for $U=20t$. At $\Delta_c\sim 19.8t$, $m$ drops to zero with a discontinuity. At the same point a discontinuity is seen the slope $\frac{\partial \delta}{\partial \Delta}$.}
\label{m1}
\end{center}
\end{figure}

{\bf{Magnetisation and staggered density}}: The staggered magnetization $m$, defined as $m = (m_A -  m_B)/2$, calculated within the renormalised mean field theory is shown in Fig.~\ref{m1}. For a given $U\gg t$, $m=0$ for $\Delta > U$ but as $\Delta$ approaches $U$, the antiferromagnetic order sets in with a jump in $m$ at $\Delta_c$. As $\Delta$ decreases further, $m$ increases approaching the saturation value. Note that for very small values of $\Delta$ where $m$ might tend to unity, our approach does not work. 

The staggered density difference $\delta=(n_A-n_B)/2$ is shown in the green curve in Fig.~\ref{m1} as a function of $\Delta$. As expected for $\Delta >U$, $\delta$ is large close to its saturation value and with decrease in $\Delta$, $\delta$ reduces monotonically for $\Delta > \Delta_c$. At $\Delta_c$,  there occurs a change in slope $\frac{\partial\delta}{\partial \Delta}$. Note that within our approach system can never attain the saturation values $m =1$ and $\delta=0$ at which the Gutzwiller factor for the spin exchange term  $g_s$ diverges and the perturbation theory fails. 
\\
\\
{\bf{Possible superconductivity in higher dimensions}}: Based on the renormalised Hamiltonian in Eq.~(\ref{H_ren}) one can see that even at half filling for the overall lattice, there is a finite hopping between $A$ and $B$ sublattices in the projected space as long as the density difference $\delta$ is non-zero. This effectively gives a doped $t-J$ model for each sublattice even at half filling. Further there are finite next nearest neighbour hopping terms within each sublattice which appear through trimer terms in the Hamiltonian in Eq.~(\ref{H_ren}). In this renormalised Hamiltonian there is a possibility that the metallic phase mentioned above can turn into a d-wave superconducting phase or $d+is$ pairing superconducting phase in higher dimensional system. The superconducting phase might survive for a larger range of $U-\Delta$ space compared to the metallic phase with support of trimer terms. This will be explored in future work. 
\\
\\
{\bf{Possible half metal phase}}: Depending upon the dimensionality of the problem and the lattice structure, it might be easier to frustrate the anti-ferromagnetic order with the help of trimer terms of Hamiltonian in Eq.~(\ref{H_ren}). The overall strength of the coefficient of various trimer terms is a non-monotonic function of $\Delta$. In the regime $\Delta\sim U$ where the staggered density difference is finite, effective next nearest neighbour hopping obtained from Fock shift decomposition of these terms might become significant and start competing with the nearest neighbour hopping term. In this case even at half-filling the effective low energy Hamiltonian does not have particle-hole symmetry and $m_A\ne -m_B$. Rather than having just non zero staggered magnetization $m=(m_A-m_B)/2$, there might be a non zero uniform magnetisation $m_F=(m_A+m_B)/2$ as well resulting in the Ferrimagnetic order for $\Delta \sim U$. Further due to different gaps in the single particle DOS for up and down spin channels, there is a possibility that the half-metallic phase appears as a precursor to the metallic phase mentioned above. A similar mechanism for half-metal has been seen  in the doped IHM~\cite{AG2} for weak to intermediate strength of $U$ and $\Delta$ where particle hole symmetry is broken explicitly by adding holes into the system while in the extremely correlated limit presented in this work, trimer terms can break the particle hole symmetry spontaneously. 
\\
\\
{\bf{Non-monotonic behaviour of Neel temperature with $\Delta$}}: The renormalised Hamiltonian in Eq.~(\ref{H_ren}) is illuminating enough to predict the behaviour of the Neel temperature for the AFM order in the IHM in large $U$ and $\Delta$ regime at half-filling. For $U\gg t$ but $\Delta\sim t$, the IHM maps to the modified $t-J$ model with an additional ionic potential term and with spin-exchange term given by $\tilde{J} = 4t^2U/(U^2-\Delta^2)$ ~\cite{Soumen}. Note that in this limit doublons are projected out from the low energy Hilbert space from all sites. In this case the Neel temperature of the AFM order should obey $\tilde{J}$ and hence increase as $\Delta$ increases. In fact this was observed in DMFT+CTQMC calculation for the IHM at half filling ~\cite{Soumen} where it was shown that for $U$ as high as $16t$, up to $\Delta$ little less than $U$, $T_N \sim \tilde{J}/4$~\cite{footnote}. But for $\Delta \ge U$  a sudden drop in $T_N$ was observed which could not be explained based on the spin exchange coupling $\tilde{J}$.

Our current renormalised Hamiltonian sheds light on this non-monotonic behaviour of $T_N$ since it is valid for $U\sim \Delta$ as well as for $\Delta >U$ regime. In this regime the coefficient of spin exchange term is $\tilde{\tilde{J}}=2t^2/(U+\Delta)$ which decreases with increase in $\Delta$. 
Hence for $U \gg t$, for small values of $\Delta \le U$ $T_N$ follows $\tilde{J}$  and hence $T_N$ increases with $\Delta$. As $\Delta$ increases further $T_N$ starts to follow the new coupling $\tilde{\tilde{J}}$ and starts decreasing with increase in $\Delta$. 

To summarise, in the strongly correlated limit of the ionic Hubbard model, interplay of $U$ and $\Delta$ promises a rich phase diagram and our formalism of renormalised Hamiltonian obtained by Gutzwiller projection of holes on one sublattice and doublons on another sublattice further implemented by Gutzwiller approximation is illuminating enough to give insight into this exotic physics. 
\section{Strongly Correlated Binary Alloys}
In this section we will discuss the physics of hole projection in context of strongly correlated limit of binary allows, modelled with the Hubbard model in the presence of a binary disorder. The Hamiltonian for this system is
\be
\mathcal{H}=-t\sum_{<ij>}c_{i\sigma}^{\dagger}c_{j\sigma} + U\sum_{i}n_{i\uparrow}n_{i\downarrow}-\sum_{i}(\mu-\epsilon_{i})n_{i}
\label{binaryH}
\ee

where $\epsilon_{i}$ is the random onsite energy drawn from the probability distribution function
\be
p_{\epsilon}(\epsilon_{i})=x\delta\bigg(\epsilon_{i}+\dfrac{V}{2}\bigg)+(1-x)\delta\bigg(\epsilon_{i}-\dfrac{V}{2}\bigg)
\label{prob}
\ee
Here, $x$ and $1-x$ are the fractions of the lattice sites with energies $-\dfrac{V}{2}$ and $\dfrac{V}{2}$ respectively. We label sites with $\epsilon(i)=-V/2$ as $A$ sites and sites with $\epsilon(i)=V/2$ as B sites. At half-filling, the above Hamiltonian is particle hole symmetric only if the percentages of $A$ and $B$ sites are equal.

Most of the earlier studies have solved this model using variants of DMFT in weak to intermediate limit of $U/t$~\cite{Hofstetter1, Byczuk1,Byczuk2,Hassan}. Using DMFT+QMC, this model has also been solved at finite temperature in the limit of sufficiently large $U$ and $V$~\cite{Byczuk2}. We are interested in strongly correlated strongly disordered limit of this model, that is, $U\sim V \gg t$.  The single site energetics is similar to IHM, that is, holes are projected out from Hilbert space at A sites and doublons are projected out from Hilbert space at B sites. The difference here is that the hole projected sites and doublon projected sites are randomly distributed on the lattice in each disorder configuration.  This makes all three type of  nearest neighbour bonds possible: AA, BB  and AB. Also in three site processes, as we will show later, there are many more hopping processes possible which do not occur for IHM. Every disorder configuration has a different combination of two site and three site hopping terms due to different environment of a site in each configuration. 

\subsection{Similarity transformation}
The nearest neighbour hopping processes between two sites can be classified as follows depending upon which sites are involved in the hopping; $AA$ sites, $BB$ sites or $A,B$ sites and whether the hopping process changes the number of doublons or not.
\[
{H_{t}}^{AA}={H_{t}^{+}}_{A\rightarrow A} +{H_{t}^{-}}_{A\rightarrow A}+{H_{t}^{0}}_{A\rightarrow A}
\]
\[
{H_{t}}^{BB}={H_{t}^{+}}_{B\rightarrow B} +{H_{t}^{-}}_{B\rightarrow B}+{H_{t}^{0}}_{B\rightarrow B}
\]
\bea
{H_{t}}^{AB}={H_{t}^{+}}_{A\rightarrow B} + {H_{t}^{+}}_{B\rightarrow A}  +{H_{t}^{-}}_{A\rightarrow B}+{H_{t}^{-}}_{B\rightarrow A}\nonumber \\
+{H_{t}^{0}}_{A\rightarrow B}+{H_{t}^{0}}_{B\rightarrow A}
\eea

Since a $A$ type site has doublons allowed in the low energy sectors and holes should be projected out while on B type sites the reverse happens, one needs to do different similarity transformations on the local Hamiltonian depending on whether the bond is $AA$ type, $BB$ type or $AB$ type. 
\[iS^{AA}=\dfrac{1}{U}({H_{t}^{+}}_{A\rightarrow A}-{H_{t}^{-}}_{A \rightarrow A}) \] 
\[iS^{BB}=\dfrac{1}{U}({H_{t}^{+}}_{B\rightarrow B}-{H_{t}^{-}}_{B \rightarrow B}) \] 
\bea
iS^{AB}=\dfrac{1}{U+V}({H_{t}^{+}}_{A \rightarrow B}-{H_{t}^{-}}_{B \rightarrow A})\nonumber \\
+\dfrac{1}{V}({H_{t}^{0}}_{A\rightarrow B}-{H_{t}^{0}}_{B\rightarrow A}) 
\eea

Note that $S^{AA}$ and $S^{BB}$ are perturbative in $t/U$ while $S^{AB}$ has term which are perturbative in $t/(U+V)$ or $t/V$.

If we consider the commutators of  the type $[S^{\alpha\beta},H_{t}^{\alpha\beta}]$ and $[S^{\alpha\beta},[S^{\alpha\beta},H_{0}^{\alpha\beta}]]$, we get terms which connect the low energy sector to the high energy sector which must be removed through suitable similarity transformation. The terms that do not interconnect the low energy sector to the high energy sector constitute the  effective Hamiltonian. Effective Hamiltonian itself is a function of 
disorder configuration. In a disorder configuration, dimer terms in $H_{eff}$ depends on whether bonds are $AA$, $BB$ or $AB$ type. 
\bea
&\mathcal{H}_{eff}=H_{0}+{H_{t}^{0}}_{A\rightarrow A} + {H_{t}^{0}}_{B\rightarrow B} +{H_{t}^{+}}_{B\rightarrow A}+{H_{t}^{-}}_{A\rightarrow B} \nonumber \\ &+\dfrac{1}{U}[{H_{t}^{+}}_{A\rightarrow A},{H_{t}^{-}}_{A\rightarrow A}] +
\dfrac{1}{U}[{H_{t}^{+}}_{B\rightarrow B},{H_{t}^{-}}_{B\rightarrow B}] \nonumber \\ 
&+\dfrac{1}{U+V}[{H_{t}^{+}}_{A\rightarrow B},{H_{t}^{-}}_{B\rightarrow A}]+\dfrac{1}{V}[{H_{t}^{0}}_{A\rightarrow B},{H_{t}^{0}}_{B\rightarrow A}] \nonumber \\
&+\dfrac{1}{2}\bigg(\dfrac{1}{U}+\dfrac{1}{V}\bigg)\big([{H_{t}^{+}}_{A\rightarrow A}+{H_t^{-}}_{B\rightarrow B},{H_{t}^{0}}_{B\rightarrow A}]\big)  \nonumber \\
&-\dfrac{1}{2}\bigg(\dfrac{1}{U}+\dfrac{1}{V}\bigg)\big([{H_{t}^{-}}_{A\rightarrow A}+{H_t^{-}}_{B\rightarrow B},{H_{t}^{0}}_{A\rightarrow B}]\big)
\label{Heff_binary}
\eea
\subsection{Effective Low energy Hamiltonian in terms of projected fermions}
Now we represent the effective low energy Hamiltonian of Eq.~(\ref{Heff_binary}) in terms of projected fermionic operators on A and B sites as defined in Eq.~(\ref{cnewA})  and(\ref{cnewB}). 
Let us first consider the $\mathcal{O}(t)$ hopping terms which are confined in the low energy Hilbert space and are represented as,
\bea
H_{1,low}^{A_i,A_j}=H_{t A\rightarrow A}^{0}(i,j) = -t\sum_\sigma[\tilde{c}_{iA\sigma}^{\dagger}\tilde{c}_{jA\sigma}+h.c.] \nonumber \\
H_{1,low}^{B_i,B_j}=H_{t B \rightarrow B}^{0}(i,j)=-t\sum_\sigma[\tilde{\tilde{c}}_{iB\sigma}^{\dagger}\tilde{\tilde{c}}_{jB\sigma} + h.c.] \nonumber \\
\eea
Here, ${H_{t}^{0}}_{A\rightarrow A}$ involves hopping of a doublon while ${H_{t}^{0}}_{B\rightarrow B}$ involves hopping of a hole.
\bea
H_{1,low}^{A_i,B_j}=H_{t A\rightarrow B}^{-}(i,j)+ H_{t B\rightarrow A}^{+}(i,j) \nonumber \\
= -t\sum_\sigma [\tilde{c}_{iA\sigma}^{\dagger}\tilde{\tilde{c}}_{jB\sigma}+ h.c.]  
\label{order_t_binary}
\eea

\underline{\bf{O($t^2/U$) Dimer terms:}}\\
Now we consider $\mathcal{O}(t^2/U)$ dimer terms obtained from $\dfrac{1}{U}[{H_{t}^{+}}_{\alpha\rightarrow \alpha},{H_{t}^{-}}_{\alpha\rightarrow \alpha}]$ terms with $\alpha=A,B$. 
 Let us first look at the $AA$ term. $\dfrac{1}{U}[{H_{t}^{+}}_{A\rightarrow A},{H_{t}^{-}}_{A\rightarrow A}] \sim -\dfrac{1}{U} {H_{t}^{-}}_{A\rightarrow A}{H_{t}^{+}}_{A\rightarrow A}$ since the first term in the commutator requires a hole to start with which lies in the high energy sector for A type sites. The dimer term corresponding to this commutator is $H_{dimer}^{A_i,A_j}$,
\begin{equation}
=-\dfrac{t^2}{U}\sum_{\sigma}[{X_{iA}}^{\sigma\leftarrow \sigma}{X_{jA}}^{\bar{\sigma}\leftarrow \bar{\sigma}}-
{X_{iA}}^{\sigma\leftarrow \bar{\sigma}}{X_{jA}}^{\bar{\sigma}\leftarrow \sigma}+ j \leftrightarrow i ]
\end{equation}

This in terms of projected operators can be expressed as

\bea
\frac{J}{2}\sum_{\sigma}[\tilde{c}_{iA\bar{\sigma}}\tilde{c}_{iA\sigma}^{\dagger}\tilde{c}_{jA\sigma}\tilde{c}_{jA\bar{\sigma}}^{\dagger}- \tilde{c}_{iA\bar{\sigma}}\tilde{c}_{iA\bar{\sigma}}^{\dagger}\tilde{c}_{jA\sigma}\tilde{c}_{jA\sigma}^{\dagger}] \nonumber \\
    =J\mathcal{P}_h\bigg(S_{iA}.S_{jA}-\dfrac{(2-n_{iA})(2-n_{jA})}{4}\bigg)\mathcal{P}_h
\label{JAA}
\eea
with $J=4t^2/U$. A factor of $4= 2 \times 2$ comes from spin summation and from hoppings from $i$ to $j$ site first or vice versa.  
A similar analysis can be extended in case of B sites. $\dfrac{1}{U}[{H_{t}^{+}}_{B\rightarrow B},{H_{t}^{-}}_{B\rightarrow B}] \sim -\dfrac{1}{U} {H_{t}^{-}}_{B\rightarrow B}{H_{t}^{+}}_{B\rightarrow B}$ since the first term in the commutator requires a doublon to start with which lies in the high energy sector for B type sites. The dimer term corresponding to this commutator is $H_{dimer}^{B_i,B_j}$,

\begin{equation}
= -\dfrac{t^2}{U}\sum_{<ij>,\sigma}[{X_{iB}}^{\sigma\leftarrow \sigma}{X_{jB}}^{\bar{\sigma}\leftarrow \bar{\sigma}}-
{X_{iB}}^{\sigma\leftarrow \bar{\sigma}}{X_{jB}}^{\bar{\sigma}\leftarrow \sigma} + j \leftrightarrow i]
\end{equation}
Again, in terms of projected operators it is,
\bea
-\dfrac{J}{2}\sum_{\sigma}[\tilde{\tilde{c}}_{iB\sigma}^{\dagger}\tilde{\tilde{c}}_{iB\sigma}\tilde{\tilde{c}}_{jB\bar{\sigma}}^{\dagger}\tilde{\tilde{c}}_{jB\bar{\sigma}}-\tilde{\tilde{c}}_{iB\sigma}^{\dagger}\tilde{\tilde{c}}_{iB\bar{\sigma}}\tilde{\tilde{c}}_{jB\bar{\sigma}}^{\dagger}\tilde{\tilde{c}}_{jB\sigma}]\nonumber \\
    =J\mathcal{P}_{d}\bigg(S_{iB}.S_{jB}-\dfrac{n_{iB}n_{jB}}{4}\bigg)\mathcal{P}_{d}
\label{JBB}
\eea

There are also $t^2/(U+V)$ order terms obtained from hopping of a spin-1/2 from site $A$ to $B$ and back. In $H_{eff}$ the corresponding term for this process is $\dfrac{1}{U+V}[{H_{t}^{+}}_{A\rightarrow B},{H_{t}^{-}}_{B\rightarrow A}]$ which, as explained in section on IHM, can be expressed as
\be 
H_{dimer}^{A_i,B_j}=J_1(S_{iA}.S_{jB}-(2-\hat{n}_{iA})\hat{n}_{jB}/4)
\label{JAB}
\ee
with $J_2=\f{2t^2}{U+V}$. Note that all above expressions are defined in projected Hilbert space.
 
The dimer term corresponding to   $[{H_{t}^{0}}_{A\rightarrow B},{H_{t}^{0}}_{B\rightarrow A}]$  involves hopping of a particle or a doublon from one site to the nearest neighbour site and back to the initial site as shown in Fig.~[\ref{dimer2}]. This process is of order $t^2/V$ and the corresponding expression is given in Eq.~(\ref{hd_hopp}). 
\\
 \begin{figure}[h]
    \centering
    \includegraphics[width=8cm]{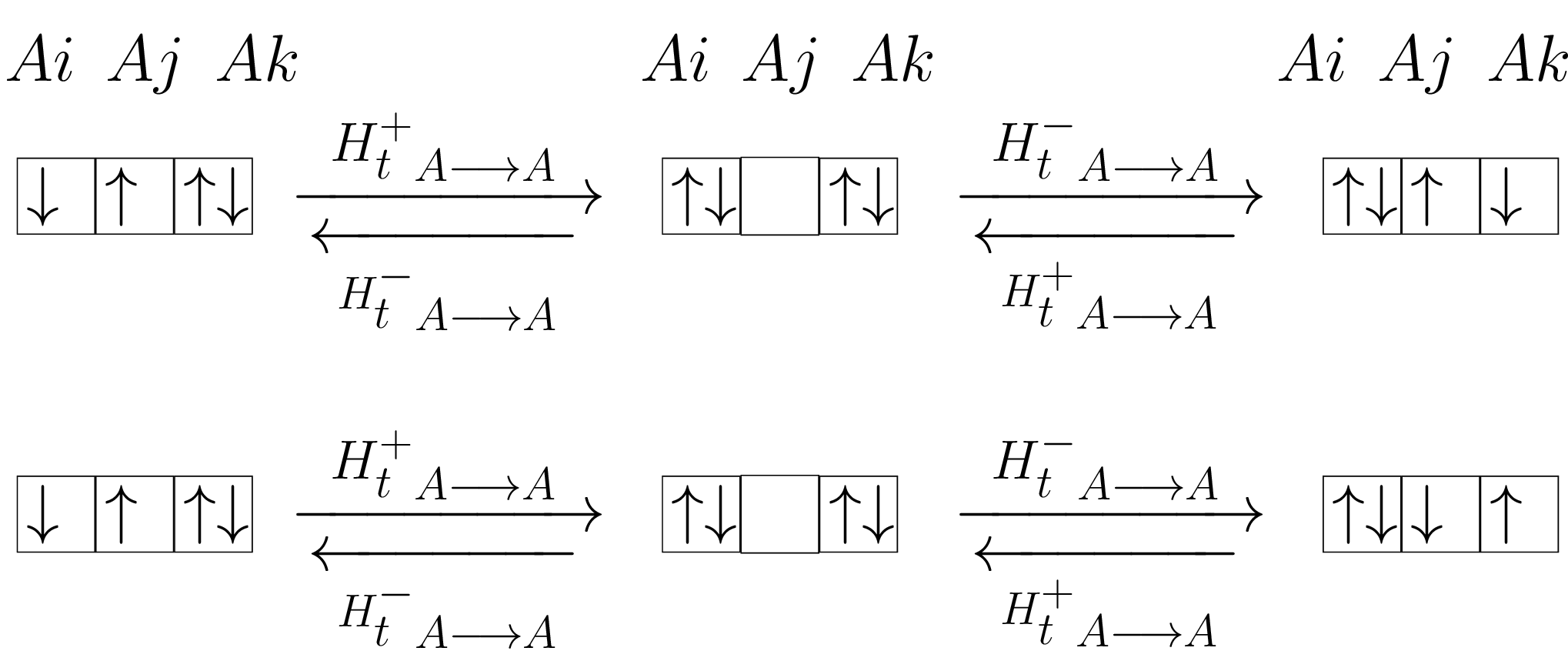}
    \caption{Trimer term on $AAA$ sites for correlated binary alloy model. }
	\label{aaa} 
   \end{figure}
\underline{\bf{O($t^2/U$) Trimer terms}}:\\
Since on each site there is possibility of having an $A$ type site or $B$ type site, in total there are $8$ trimer terms possible arising from various commutators in $H_{eff}$. Trimer terms from the commutator involving only $A$ type sites $\dfrac{1}{U}[{H_{t}^{+}}_{A\rightarrow A},{H_{t}^{-}}_{A\rightarrow A}]$ involves hopping of a particle from the intermediate  site resulting in the formation of a doublon in the nearest neighbour site and the other doublon unpairs in two ways : one in the spin preserving way, other in the spin flip way as shown in Fig.~[\ref{aaa}]. Eventually we get $H_{trimer}^{AAA}(i,j,k)$ as 

\[
-\dfrac{t^2}{U}\sum_\sigma[X_{jA}^{\sigma\leftarrow 0}X_{kA}^{\bar{\sigma} \leftarrow d}X_{iA}^{d \leftarrow \bar{\sigma}}X_{jA}^{0\leftarrow \sigma} + h.c.]\]
\[
+\f{t^2}{U}\sum_\sigma[X_{jA}^{\sigma\leftarrow 0}X_{kA}^{\bar{\sigma} \leftarrow d}X_{iA}^{d \leftarrow \sigma}X_{jA}^{0\leftarrow \bar{\sigma}}+h.c.]\]
\[
=\dfrac{t^2}{U}\sum_{\sigma}[\tilde{c}_{iA\sigma}^{\dagger}\tilde{c}_{jA\bar{\sigma}}\tilde{c}_{jA\bar{\sigma}}^{\dagger}\tilde{c}_{kA\sigma}-\tilde{c}_{iA\bar{\sigma}}^{\dagger}\tilde{c}_{jA\bar{\sigma}}\tilde{c}_{jA\sigma}^{\dagger}\tilde{c}_{kA\sigma}] + h.c. \]
\be
=\dfrac{t^2}{U}\sum_{\sigma}\mathcal{P}_{h}(c_{iA\sigma}^{\dagger}(1-\hat{n}_{jA\bar{\sigma}})c_{kA\sigma}+c_{iA\bar{\sigma}}^{\dagger}c_{jA\sigma}^{\dagger}c_{jA\bar{\sigma}}c_{kA\sigma}+h.c.)\mathcal{P}_{h}
\label{AAAeqn}
\ee

A similar trimer term on $BBB$ sites is obtained from  $\dfrac{1}{U}[{H_{t}^{+}}_{B\rightarrow B},{H_{t}^{-}}_{B\rightarrow B}]$.  In the BBB trimer terms, effective next nearest neighbour hopping of hole takes place just like in $AAA$ terms it is the effective next nearest neighbour hopping of a doublon which takes place. The corresponding trimer term can be expressed as $H_{trimer}^{BBB}$ 
\[
=-\dfrac{t^2}{U}\sum_{\sigma}[X_{iB}^{\sigma\leftarrow0}X_{jB}^{\bar{\sigma} \leftarrow d}X_{jB}^{d \leftarrow \bar{\sigma}}X_{kB}^{0\leftarrow \sigma} + h.c.] \]
\[
+\dfrac{t^2}{U}\sum_\sigma[X_{iB}^{\bar{\sigma}\leftarrow0}X_{jB}^{\sigma \leftarrow d}X_{jB}^{d \leftarrow \bar{\sigma}}X_{kB}^{0\leftarrow \sigma}+h.c.] \]

\[=-\dfrac{t^2}{U}\sum_{\sigma}[\tilde{\tilde{c}}_{iB\sigma}^{\dagger}\tilde{\tilde{c}}_{jB\bar{\sigma}}^{\dagger}\tilde{\tilde{c}}_{jB\bar{\sigma}}\tilde{\tilde{c}}_{kB\sigma}-\tilde{\tilde{c}}_{iB\bar{\sigma}}^{\dagger}\tilde{\tilde{c}}_{jB\sigma}^{\dagger}\tilde{\tilde{c}}_{jB\bar{\sigma}}\tilde{\tilde{c}}_{kB\sigma}]\]
\be
=-\dfrac{t^2}{U}\sum_{\sigma}\mathcal{P}_{d}(c_{iB\sigma}^{\dagger}n_{jB\bar{\sigma}}c_{kB\sigma}-c_{iB\bar{\sigma}}^{\dagger}c_{jB\sigma}^{\dagger}c_{jB\bar{\sigma}}c_{kB\sigma}+h.c.)\mathcal{P}_{d}
\label{BBBeqn}
\end{equation}

Then there are $ABA$ and $BAB$ type trimer terms, which are of order $t^2/V$. Note that similar terms also appeared in IHM and are represented in Fig.~[\ref{trimer1}] and Fig.~[\ref{trimer2}]. 
Below we summarise their forms for the case of binary alloy model
\bea
H_{trimer}^{Ai,Bj,Ak} =-\frac{t^2}{V}\sum_\sigma\mathcal{P}(c^{\dagger}_{kA\sigma}[n_{jB\bar{\sigma}}c_{iA\sigma} \nonumber \\
- c_{iA\bar{\sigma}}c^{\dagger}_{jB\bar{\sigma}}c_{jB\sigma}])\mathcal{P}
\eea
\bea
H_{trimer}^{Bi,Aj,Bk}= -\frac{t^2}{V}\sum_{\sigma}\mathcal{P}(c_{kB\sigma}[(1-n_{iA\bar{\sigma}})c_{jB\sigma}^{\dagger} \nonumber \\
+c_{iA\sigma}^{\dagger}c_{iA\bar{\sigma}}c_{jB\bar{\sigma}}^{\dagger}])\mathcal{P}
\eea 

\underline{\bf{AAB and BBA trimer terms}}:\\
Next we consider the remaining trimer terms, namely, $AAB (or BAA)$ and $BBA$(or $ABB$) type terms. We would like to emphasize that these terms never appear in strongly correlated limit of IHM presented in earlier section and are characteristic of random arrangement of A and B type sites in binary alloy model. 
 
\begin{figure}[ht]
    \centering
    \includegraphics[width=8cm]{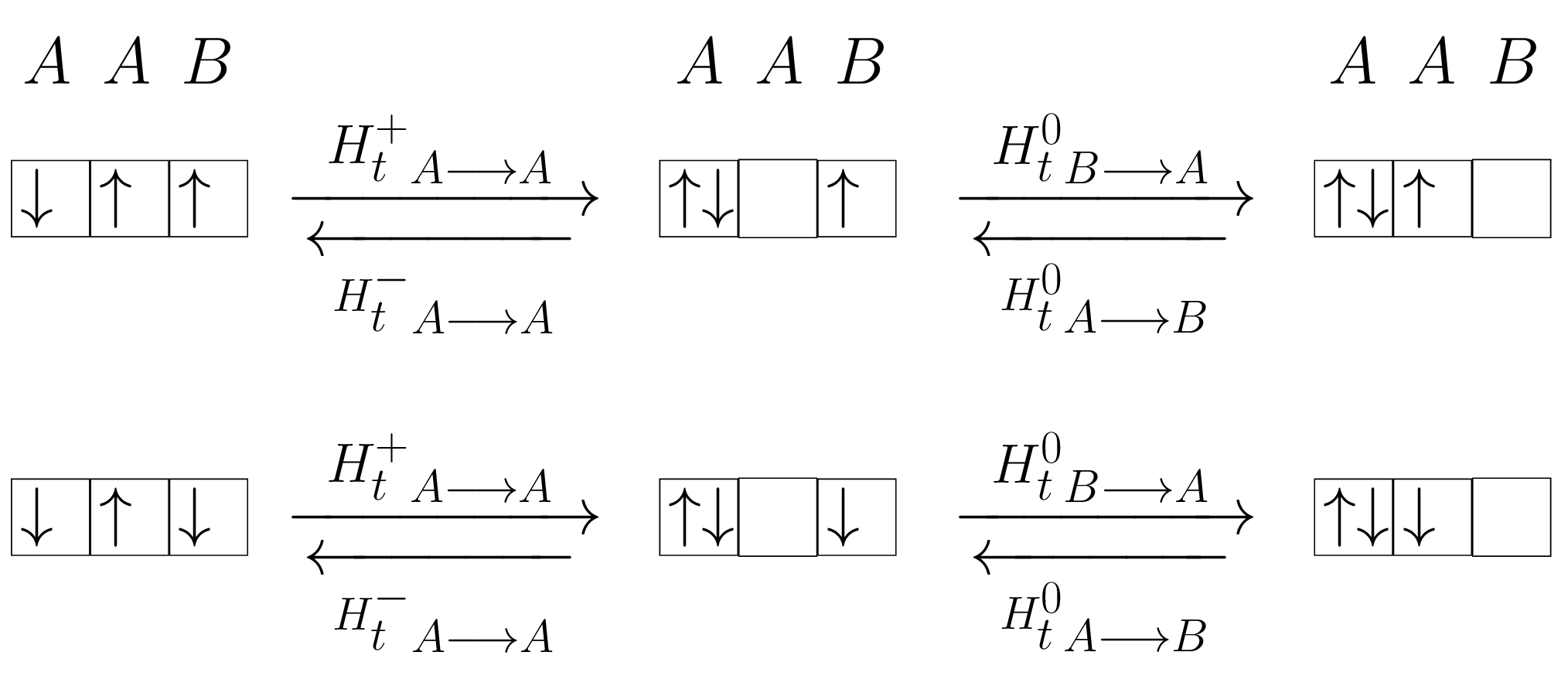}
    \caption{AAB Trimer processes for correlated binary alloy model.}
\label{AAB}
    \end{figure}

The AAB trimer terms, shown in Fig.~[\ref{AAB}], arise from the commutator $\f{t^2(U+V)}{2UV}[{H_{t}^{+}}_{A\rightarrow A},{H_{t}^{0}}_{B\rightarrow A}]\sim -\f{K}{t^2}{H_{t}^{0}}_{B\rightarrow A}{H_{t}^{+}}_{A\rightarrow A}$ where we have define the coupling strength for this term $K=\f{t^2(U+V)}{2UV}$. 
This is because the first term of the commutator requires a hole at the intermediate A site to begin with which is energetically unfavourable. As shown in Fig.~[\ref{AAB}], consists of  usual spin preserving and spin flip terms. In one case, the spin at the intermediate site remains the same as the initial state and in the other case it flips.

The fermionic representation of these terms $H_{trimer}^{A_i,A_k,B_j}$ is as follows,
\[
-K\sum_\sigma\eta(\sigma)[X_{kA}^{\sigma\leftarrow 0}X_{jB}^{0\leftarrow\sigma}X_{iA}^{d\leftarrow \bar{\sigma}}X_{kA}^{0\leftarrow \sigma} \]
\vskip-0.3cm
\[
+X_{kA}^{\bar{\sigma}\leftarrow 0}X_{jB}^{0\leftarrow \bar{\sigma}}X_{iA}^{d\leftarrow \bar{\sigma}}X_{kA}^{0\leftarrow \sigma}]\]
\vskip-0.3cm
\[
=K\sum_{\sigma}(\tilde{c}_{iA\sigma}^{\dagger}\tilde{c}_{kA\bar{\sigma}}\tilde{c}_{kA\bar{\sigma}}^{\dagger}\tilde{\tilde{c}}_{jB\sigma}-\tilde{c}_{iA\sigma}^{\dagger}\tilde{c}_{kA\sigma}\tilde{c}_{kA\bar{\sigma}}^\dagger\tilde{\tilde{c}}_{jB\bar{\sigma}})
\]
\vskip-0.3cm
\be
=K\sum_{\sigma}\mathcal{P}(c_{iA\sigma}^{\dagger}(1-n_{kA\bar{\sigma}})c_{jB\sigma}+c_{iA\sigma}^{\dagger}c_{kA\bar{\sigma}}^{\dagger}c_{kA\sigma}c_{jB\bar{\sigma}})\mathcal{P}
\label{AABeqn}
\ee

\begin{figure}[h]
    \centering
    \includegraphics[width=8cm]{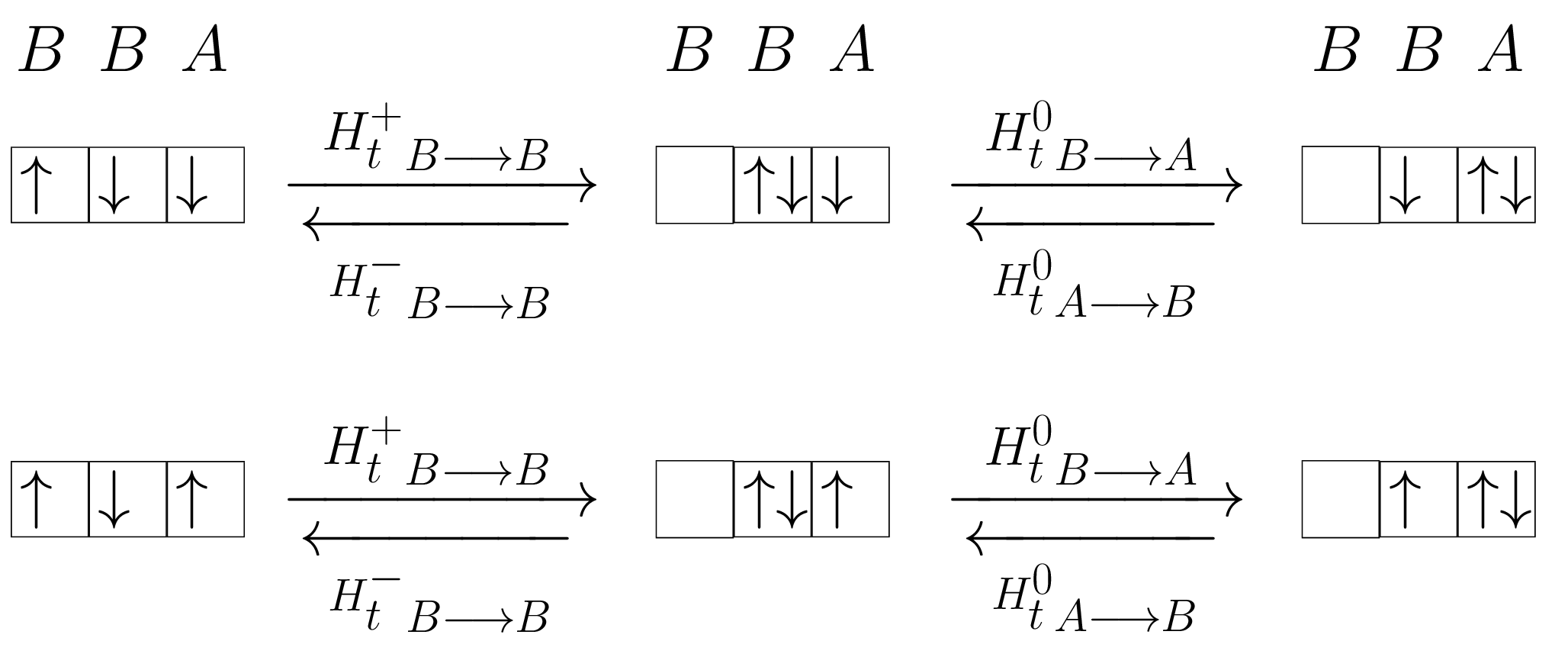}
    \caption{BBA Trimer processes for correlated binary alloy model.}
\label{BBA}
\end{figure}

Similarly, the BBA trimer terms appear from the commutator $K[{H_{t}^{+}}_{B\rightarrow B},{H_{t}^{0}}_{B\rightarrow A}]\sim -\f{K}{t^2}{H_{t}^{0}}_{B\rightarrow A}{H_{t}^{+}}_{B\rightarrow B}$. The first term in the commutator requires a doublon at the intermediate site B to start with which is energetically unfavourable. As shown in Fig.~[\ref{BBA}], These terms also come in two variants, spin preserving and spin flip at the intermediate site.

Below we represent them in terms of $X$ operators and then in terms of projected operators $H_{trimer}^{B_j,B_l,A_i}$ as 
\[
-K\sum_{\sigma}\eta(\sigma)[X_{iA}^{d\leftarrow\bar{\sigma}}X_{lB}^{\bar{\sigma}\leftarrow d}X_{lB}^{d\leftarrow\bar{\sigma}}X_{jB}^{0\leftarrow \sigma} +\]
\vskip-0.5cm
\[
X_{iA}^{d\leftarrow \sigma}X_{lB}^{\sigma\leftarrow d}X_{lB}^{d\leftarrow\bar{\sigma}}X_{jB}^{0\leftarrow \sigma}]\]
\vskip-0.5cm
\[
=-K\sum_{\sigma}(\tilde{c}_{iA\sigma}^{\dagger}\tilde{\tilde{c}}_{lB\bar{\sigma}}^{\dagger}\tilde{\tilde{c}}_{lB\bar{\sigma}}\tilde{\tilde{c}}_{jB\sigma}-\tilde{c}_{iA\sigma}^{\dagger}\tilde{\tilde{c}}_{lB\bar{\sigma}}^{\dagger}\tilde{\tilde{c}}_{lB\sigma}\tilde{\tilde{c}}_{jB\bar{\sigma}})\]
\vskip-0.5cm
\be
=-K\sum_{\sigma}\mathcal{P}(c_{iA\sigma}^{\dagger}n_{lB\bar{\sigma}}c_{jB\sigma}-c_{iA\sigma}^{\dagger}c_{lB\bar{\sigma}}^{\dagger}c_{lB\sigma}c_{jB\bar{\sigma}})\mathcal{P}
\label{BBAeqn}
\ee

The terms from the commutators $[{H_{t}^{-}}_{A\rightarrow A},{H_{t}^{0}}_{A\rightarrow B}]$   and $[{H_{t}^{-}}_{B\rightarrow B},{H_{t}^{0}}_{A\rightarrow B}] $ are the hermitian conjugate terms of the trimer terms in Eq. (\ref{AABeqn}) and (\ref{BBAeqn}) and are represented by the lower arrows in Fig.~[\ref{AAB}] and [\ref{BBA}].

\vspace{0.2cm}

\begin{figure*}[ht]
\hfill
\hspace{-3cm}
\subfigure{\frame{\includegraphics[width=6cm]{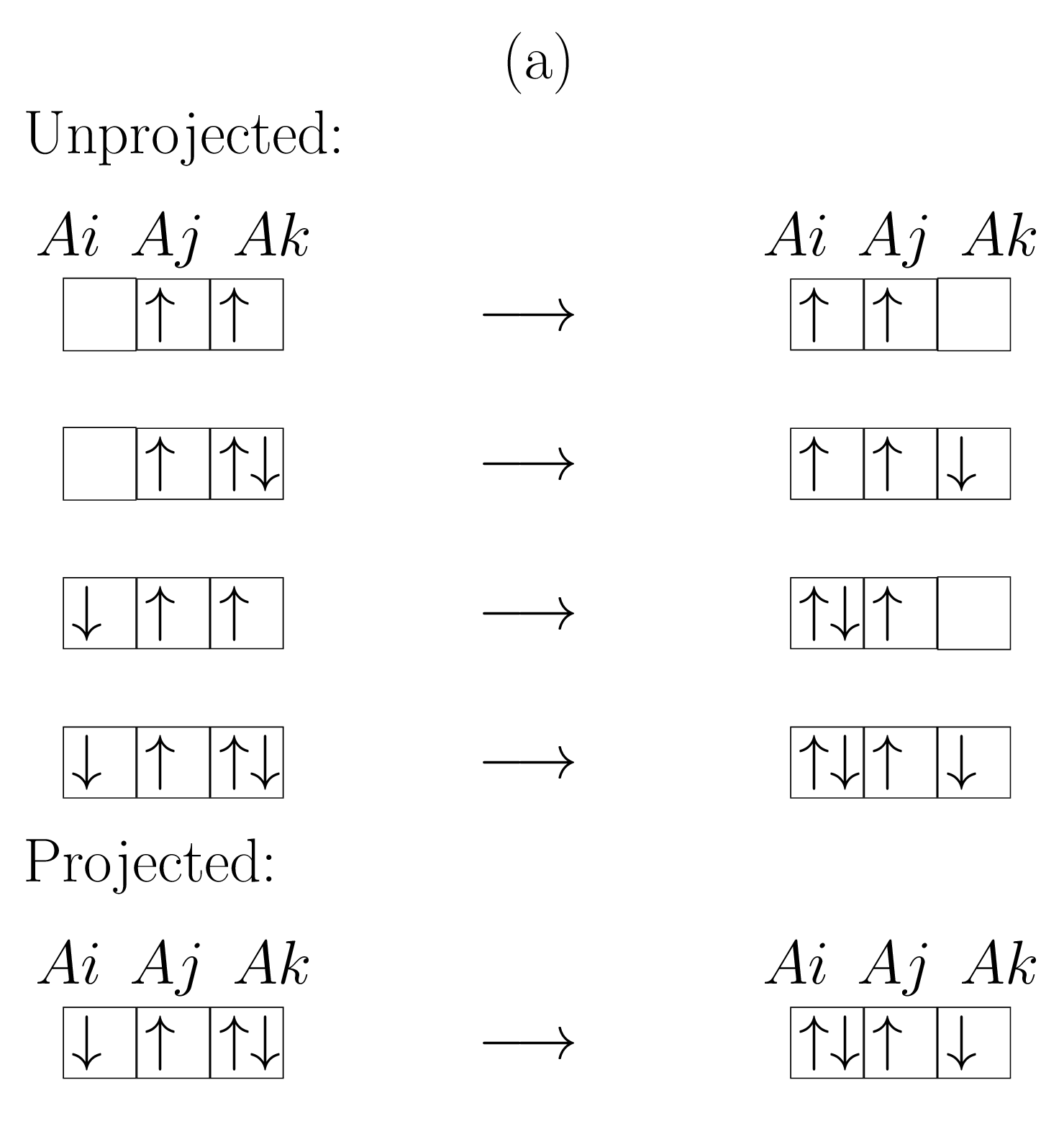}}}
\hfill
\subfigure{\frame{\includegraphics[width=6cm]{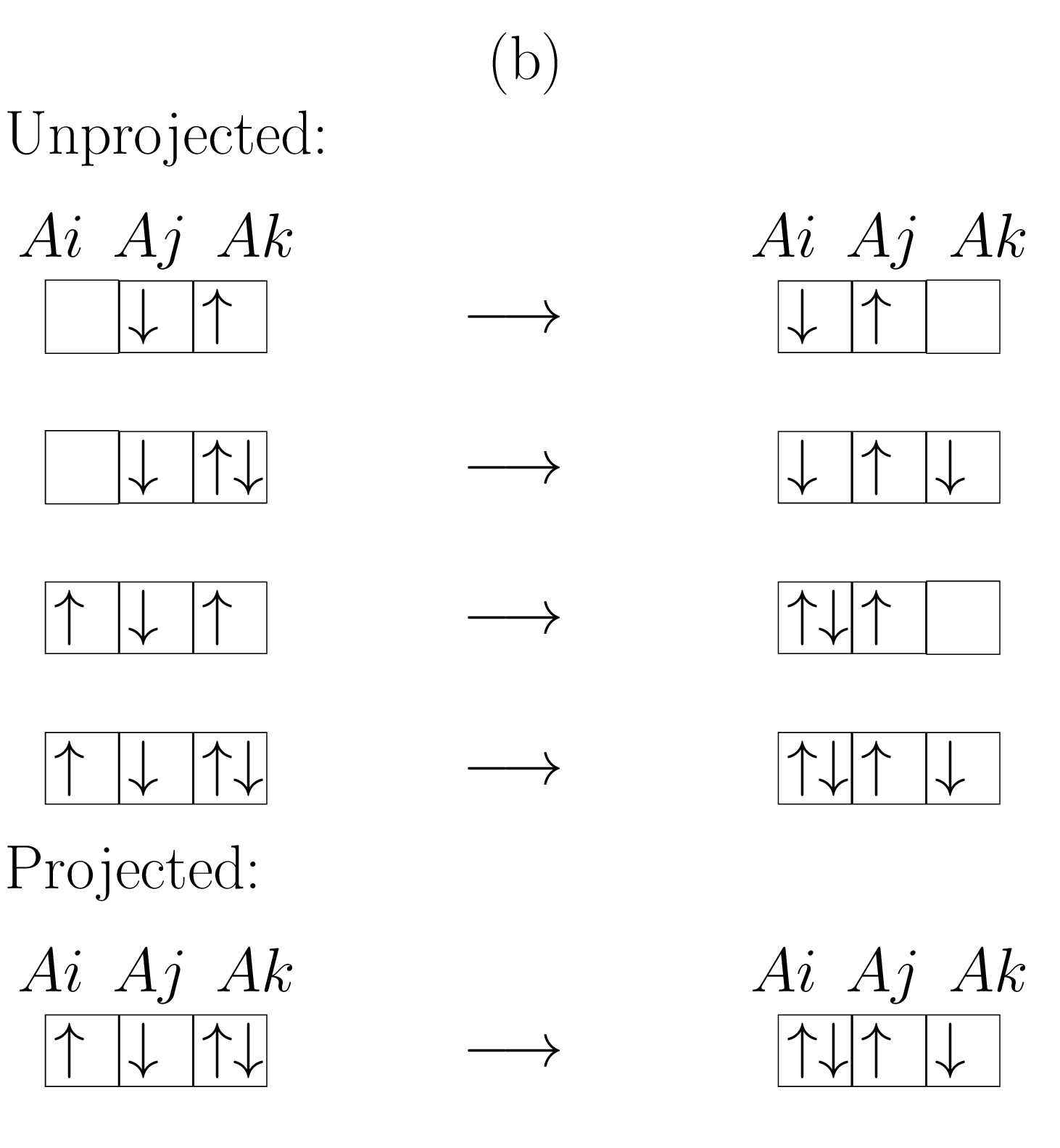}}}
\hfill
\caption{(a) Processes involved in the calculation of $g_{1\sigma}$ and $g_{2\sigma}$ which are renormalization Gutzwiller factors for AAA trimer terms.} 
\label{AAAGutz}
\end{figure*}

\subsection{Gutzwiller Approximation}
After finding various terms in the low energy effective Hamiltonian for strongly correlated binary disorder model, we will now evaluate Gutzwiller factors for various terms in $H_{eff}$ of Eq.~(\ref{Heff_binary}). The low energy effective Hamiltonian for binary alloys consists of certain dimer and trimer terms and for some of these terms we have already found the Gutzwiller factors in the section on IHM. However here, unlike in IHM, the densities on A or B sites are not homogeneous. They are site dependent and depends on the local environment. Let us first consider the hopping process of $O(t)$ between two neighbouring sites. Within Gutzwiller approximation 
\bea
H_{1,low}^{A_i,A_j} = -t\sum_\sigma \mathcal{P}_h[c^\dagger_{iA\sigma}c_{jA\sigma}+h.c.]\mathcal{P}_h \nonumber \\
= -t\sum_\sigma g_{t\sigma}^{AA}(i,j)[c^\dagger_{iA\sigma}c_{jA\sigma}+h.c.] \nonumber \\
H_{1,low}^{B_i,B_j} = -t\sum_\sigma \mathcal{P}_d[c^\dagger_{iB\sigma}c_{jB\sigma}+h.c.]\mathcal{P}_d \nonumber \\
= -t\sum_\sigma g_{t\sigma}^{BB}(i,j)[c^\dagger_{iB\sigma}c_{jB\sigma}+h.c.] \nonumber \\
H_{1,low}^{A_i,B_j} = -t\sum_\sigma \mathcal{P}[c^\dagger_{iA\sigma}c_{jB\sigma}+h.c.]\mathcal{P} \nonumber \\
= -t\sum_\sigma g_{t\sigma}^{AB}(i,j)[c^\dagger_{iA\sigma}c_{jB\sigma}+h.c.] 
\label{H1b}
\eea
As explained for $AB$ terms in section on IHM, one can evaluate these Gutzwiller factors by evaluating probability for hopping process on corresponding bonds within the projected and unprojected Hilbert space. By doing an exercise similar to the one explained in the section on IHM, we obtain, 
\bea
 g_{t\sigma}^{AA}(i,j)=\sqrt{\dfrac{({\bf{n}}_{iA}-1)({\bf{n}}_{jA}-1)}{{\bf{n}}_{iA\sigma}{\bf{n}}_{jA\sigma}}} \nonumber \\
g_{t\sigma}^{BB}(i,j)=\sqrt{\dfrac{(1-{\bf{n}}_{iB})(1-{\bf{n}}_{jB})}{(1-{\bf{n}}_{iB\sigma})(1-{\bf{n}}_{jB\sigma})}}\nonumber \\
g_{t\sigma}^{AB}(i,j)=\sqrt{\dfrac{({\bf{n}}_{iA}-1)(1-{\bf{n}}_{jB})}{{\bf{n}}_{iA\sigma}(1-{\bf{n}}_{jB\sigma})}}
\label{gtb}
\eea

Next let us consider the renormalization of $O(t^2/U)$ dimer terms which are also of three type depending upon the bond under consideration in a given disorder configuration. Within Gutzwiller approximation, couplings in Eq.~(\ref{JAA}),(\ref{JBB}) and (\ref{JAB}) get rescaled with the corresponding Gutzwiller factors to give
\bea
\hspace{-0.3cm}
H_{dimer}^{A_i,A_j} = Jg_s^{AA}(i,j)\bigg(S_{iA}.S_{jA}-\dfrac{(2-n_{iA})(2-n_{jA})}{4}\bigg) \nonumber \\
\hspace{-0.5cm}
H_{dimer}^{B_i,B_j}=Jg_s^{BB}(i,j)\bigg(S_{iB}.S_{jB}-\dfrac{n_{iB}n_{jB}}{4}\bigg) \nonumber \\
\hspace{-0.5cm}
H_{dimer}^{A_i,B_j} = J_2g_s^{AB}(i,j)\bigg(S_{iA}.S_{jB}-\dfrac{(2-n_{iA})n_{jB}}{4}\bigg)
\eea

The corresponding Gutzwiller factors are obtained, as explained for an AB term in the section on IHM, to be
\bea
g_s^{AA}(i,j)=\dfrac{1}{\sqrt{{\bf{n}}_{iA\uparrow}{\bf{n}}_{iA\downarrow}{\bf{n}}_{jA\uparrow}{\bf{n}}_{jA\downarrow}}} \nonumber \\
g_{s}^{BB}(i,j)=\dfrac{1}{\sqrt{(1-{\bf{n}}_{iB\uparrow})(1-{\bf{n}}_{iB\downarrow})(1-{\bf{n}}_{jB\uparrow})(1-{\bf{n}}_{jB\downarrow})}} \nonumber 
\eea
\be
g_{s}^{AB}(i,j)=\dfrac{1}{\sqrt{{\bf{n}}_{iA\uparrow}{\bf{n}}_{iA\downarrow}(1-{\bf{n}}_{jB\uparrow})(1-{\bf{n}}_{jB\downarrow})}}
\ee

In the calculation of Gutzwiller factors for the trimer terms, the intermediate step is unimportant, only the initial and final states are used to calculate the probabilities. 
Renormalised form of $AAA$ trimer term which is written in Eq.~(\ref{AAAeqn}) is given below
\bea
H_{trimer}^{A_i,A_j,A_k} = \dfrac{t^2}{U}\sum_{\sigma}(g_{1\sigma}^{AAA}(i,j,k)c_{iA\sigma}^{\dagger}(1-\hat{n}_{jA\bar{\sigma}})c_{kA\sigma} \nonumber \\
+g_{2\sigma}^{AAA}(i,j,k)c_{iA\bar{\sigma}}^{\dagger}c_{jA\sigma}^{\dagger}c_{jA\bar{\sigma}}c_{kA\sigma}+h.c.)
\eea
The processes in projected and unprojected spaces for the calculation of $g_{1\uparrow}$ is shown in Fig.~[\ref{AAAGutz}]. The probability of the process in  unprojected basis is $(1-{\bf{n}}_{iA\uparrow}){\bf{n}}_{iA\uparrow}(1-{\bf{n}}_{jA\downarrow})^{2}{\bf{n}}_{kA\uparrow}(1-{\bf{n}}_{kA\uparrow})$ and in the projected basis it is $({\bf{n}}_{iA}-1)(1-{\bf{n}}_{iA\uparrow})(1-{\bf{n}}_{jA\downarrow})^{2}({\bf{n}}_{kA}-1)(1-{\bf{n}}_{kA\uparrow})$. The Gutzwiller factor then comes out to be 
\be
g_{1\uparrow}^{AAA}(i,j,k)=\sqrt{\dfrac{({\bf{n}}_{iA}-1)({\bf{n}}_{kA}-1)}{{\bf{n}}_{iA\uparrow}{\bf{n}}_{kA\uparrow}}}
\ee 
 In Fig.~[\ref{AAAGutz}(b)], the processes in unprojected and projected spaces required for the calculation of $g_{2\sigma}$ are shown for which the total probability in the unprojected basis is $(1-{\bf{n}}_{iA\downarrow}){\bf{n}}_{jA\downarrow}(1-{\bf{n}}_{jA\uparrow}){\bf{n}}_{kA\uparrow}(1-{\bf{n}}_{kA\uparrow}){\bf{n}}_{jA\uparrow}(1-{\bf{n}}_{jA\downarrow}){\bf{n}}_{iA\downarrow}$ and in the projected basis is $(1-{\bf{n}}_{iA\downarrow})(1-{\bf{n}}_{jA\uparrow})({\bf{n}}_{kA}-1)(1-{\bf{n}}_{kA\uparrow})(1-{\bf{n}}_{jA\downarrow})({\bf{n}}_{iA}-1)$. The Gutzwiller factor then comes out to be 
\be
g_{2\sigma}^{AAA}(i,j,k)=\sqrt{\dfrac{({\bf{n}}_{kA}-1)({\bf{n}}_{iA}-1)}{{\bf{n}}_{jA\uparrow}{\bf{n}}_{jA\downarrow}{\bf{n}}_{kA\sigma}{\bf{n}}_{iA\bar{\sigma}}}}
\ee
 
Similarly, for the $BBB$ trimer terms of Eq.~(\ref{BBBeqn}) can be obtained by replacing ${\bf{n}}_{A\sigma}$ with $(1-{\bf{n}}_{B\sigma})$ and $({\bf{n}}_A-1)$ with $(1-{\bf{n}}_B)$ in above two equations.
 
Now we consider the trimer terms of $ABA$ and $BAB$ type for which we also calculated the Gutzwiller factors in section on IHM. The renormalised form of these terms under Gutzwiller approximation is
\bea
\hspace{-0.5cm}
H_{trimer}^{Ai,Bj,Ak} = -\frac{t^2}{V}\sum_\sigma c^{\dagger}_{kA\sigma}[g_{1\sigma}^{ABA}(i,j,k)n_{jB\bar{\sigma}}c_{iA\sigma} \nonumber \\
- g_{2\sigma}^{ABA}(i,j,k) c_{iA\bar{\sigma}}c^{\dagger}_{jB\bar{\sigma}}c_{jB\sigma}]
\eea
\bea
\hspace{-0.5cm}
H_{trimer}^{Bj,Ai,Bl}= -\frac{t^2}{V}\sum_{\sigma}c_{lB\sigma}[g_{1\sigma}^{BAB}(j,i,l)(1-n_{iA\bar{\sigma}})c_{jB\sigma}^{\dagger} \nonumber \\
+g_{2\sigma}^{BAB}(j,i,l) c_{iA\sigma}^{\dagger}c_{iA\bar{\sigma}}c_{jB\bar{\sigma}}^{\dagger}]
\eea
 Now we will calculate Gutzwiller factors for these trimer terms shown in Fig.[\ref{trimer1}] and Fig.~[\ref{trimer2}]. Fig.[\ref{trimer1}(a)] shows hopping of an $\ua$ electron from an A site to its next nearest neighbour A sites with a spin ($\da$) on the intermediate B site being preserved. In the unprojected basis, the probability for this process to happen is ${\bf{n}}_{iA\uparrow}{\bf{n}}_{jB\downarrow}^{2}(1-{\bf{n}}_{kA\ua})(1-{\bf{n}}_{iA\ua}){\bf{n}}_{kA\ua}$. It is to be noted that processes with either a down type particle or a doublon at the intermediate B site have been considered in the unprojected space. Like wise, the probability for the process to happen in the projected basis is $({\bf{n}}_{iA}-1)(1-{\bf{n}}_{kA\uparrow}){\bf{n}}_{B\downarrow}^{2}({\bf{n}}_{kA}-1)(1-{\bf{n}}_{iA\ua})$. Therefore, the Gutzwiller factor for this process is
\be
g_{1\uparrow}^{ABA}(i,j,k)=\sqrt{\dfrac{({\bf{n}}_{iA}-1)({\bf{n}}_{kA}-1)}{{\bf{n}}_{iA\uparrow}{\bf{n}}_{kA\ua}}}
\label{g1ABA}
\ee
Fig.~[\ref{trimer1}(b)] depicts hopping processes on A sublattice in which spin on the intermediate B site gets flipped. The probability in the unprojected basis for this process to occur is ${\bf{n}}_{iA\da}{\bf{n}}_{jB\ua}(1-{\bf{n}}_{jB\downarrow})(1-{\bf{n}}_{kA\uparrow})(1-{\bf{n}}_{iA\downarrow}){\bf{n}}_{jB\downarrow}(1-{\bf{n}}_{jB\uparrow}){\bf{n}}_{kA\uparrow})$ where as that in the projected basis is $({\bf{n}}_{iA}-1){\bf{n}}_{jB\ua}(1-{\bf{n}}_{kA\uparrow})(1-{\bf{n}}_{iA\downarrow}){\bf{n}}_{jB\downarrow}({\bf{n}}_{kA}-1)$ resulting in the Gutzwiller factor 
\be
g_{2\sigma}^{ABA}(i,j,k)=\sqrt{\dfrac{({\bf{n}}_{iA}-1)({\bf{n}}_{kA}-1)}{{\bf{n}}_{iA\bar{\sigma}}{\bf{n}}_{kA\sigma}(1-{\bf{n}}_{jB\ua})(1-{\bf{n}}_{jB\da})}}
\label{g2ABA}
\ee

\begin{figure*}[ht]
\hfill
\hspace{-3cm}
\subfigure{\frame{\includegraphics[width=6cm]{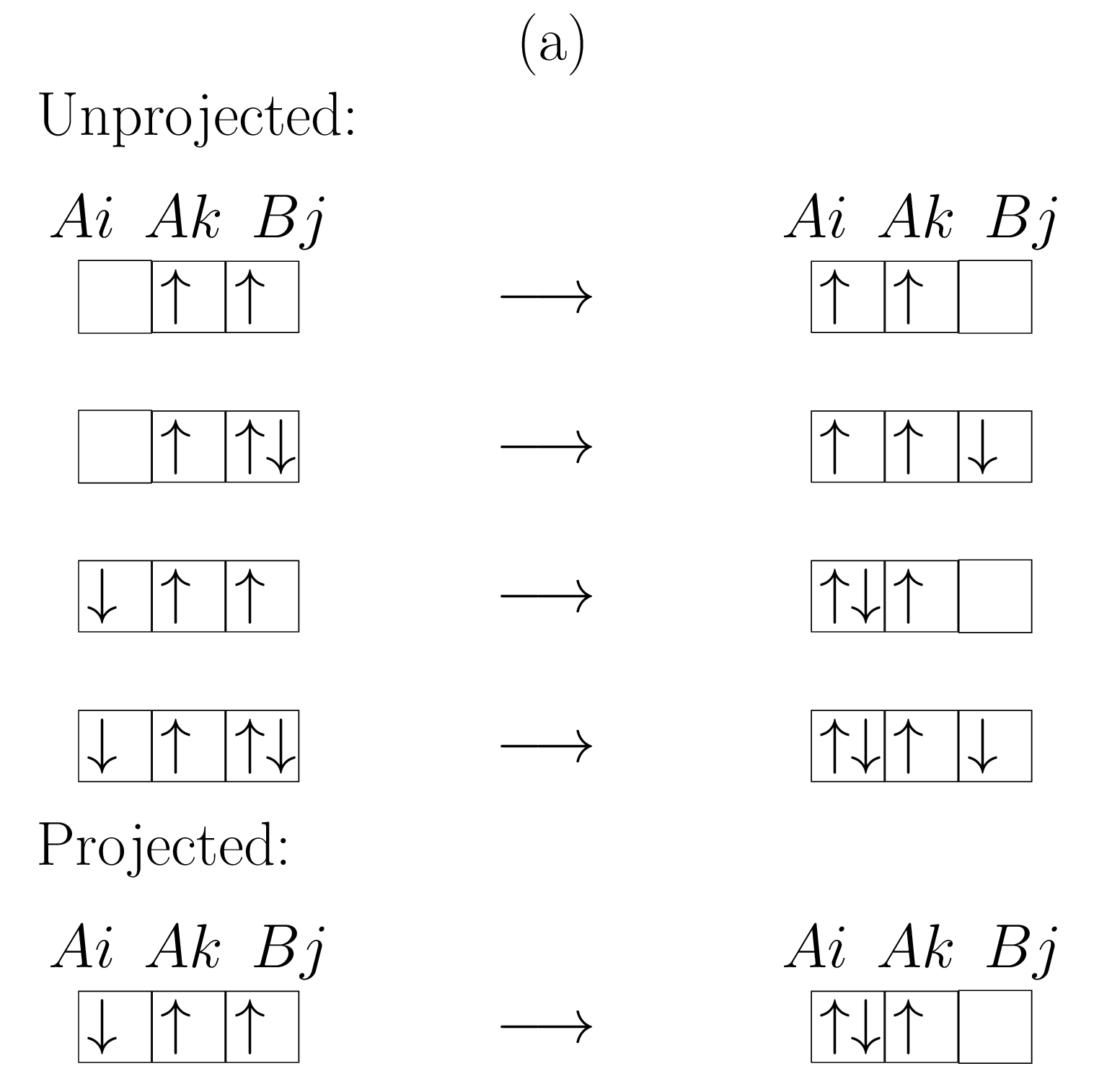}}}
\hfill
\subfigure{\frame{\includegraphics[width=6cm]{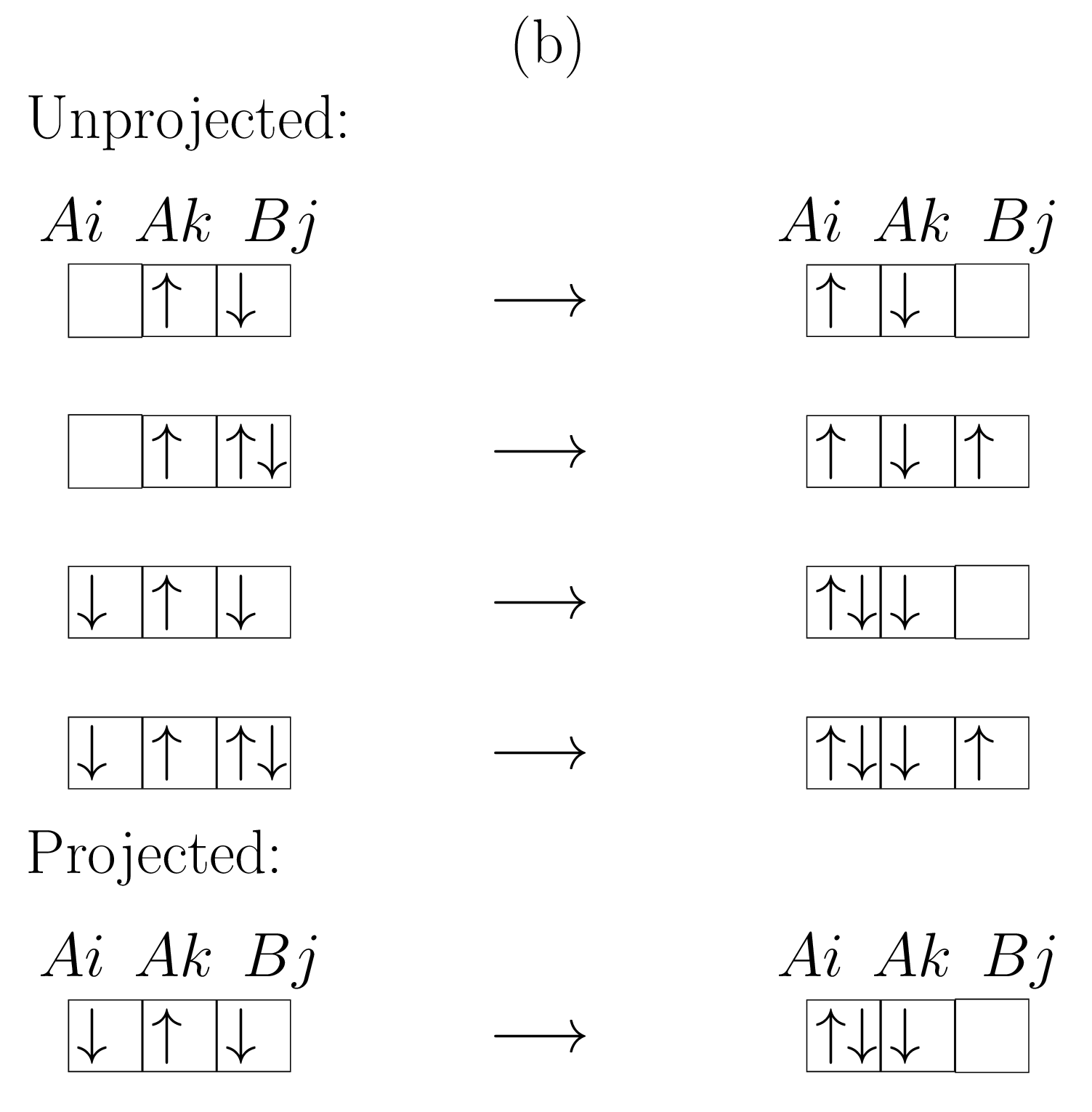}}}
\hfill
\caption{(a) Processes involved in the calculation of  $g_{1\sigma}^{AAB}$. Similar AAB physical processes with hole at intermediate A site in the unprojected basis are considered in the calculation. (b) Processes involved in calculation of $g_{2\sigma}^{AAB}$.}
\label{AABGutz}
\end{figure*}

Similarly, we can obtain the Gutzwiller factors $g_{1\sigma}^{BAB}(i,j,l)$ and $g_{2\sigma}^{BAB}(i,j,l)$ from above two equations by replacing ${\bf{n}}_{A\sigma}$ with $(1-{\bf{n}}_{B\sigma})$ and $({\bf{n}}_{A}-1)$ with $(1-{\bf{n}}_B)$. 

\begin{figure*}[ht]
\hfill
\hspace{-3cm}
\subfigure{\frame{\includegraphics[width=6cm]{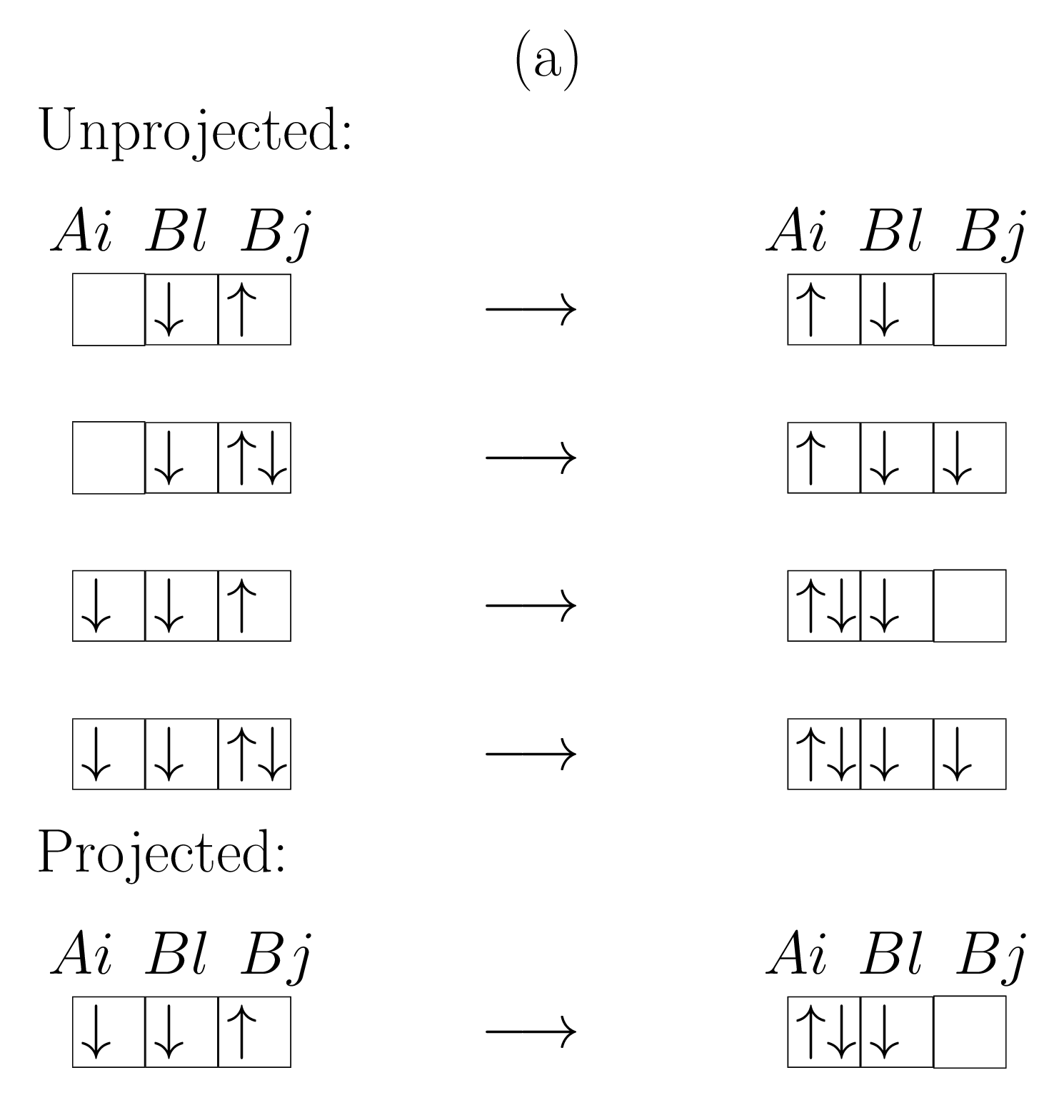}}}
\hfill
\subfigure{\frame{\includegraphics[width=6cm]{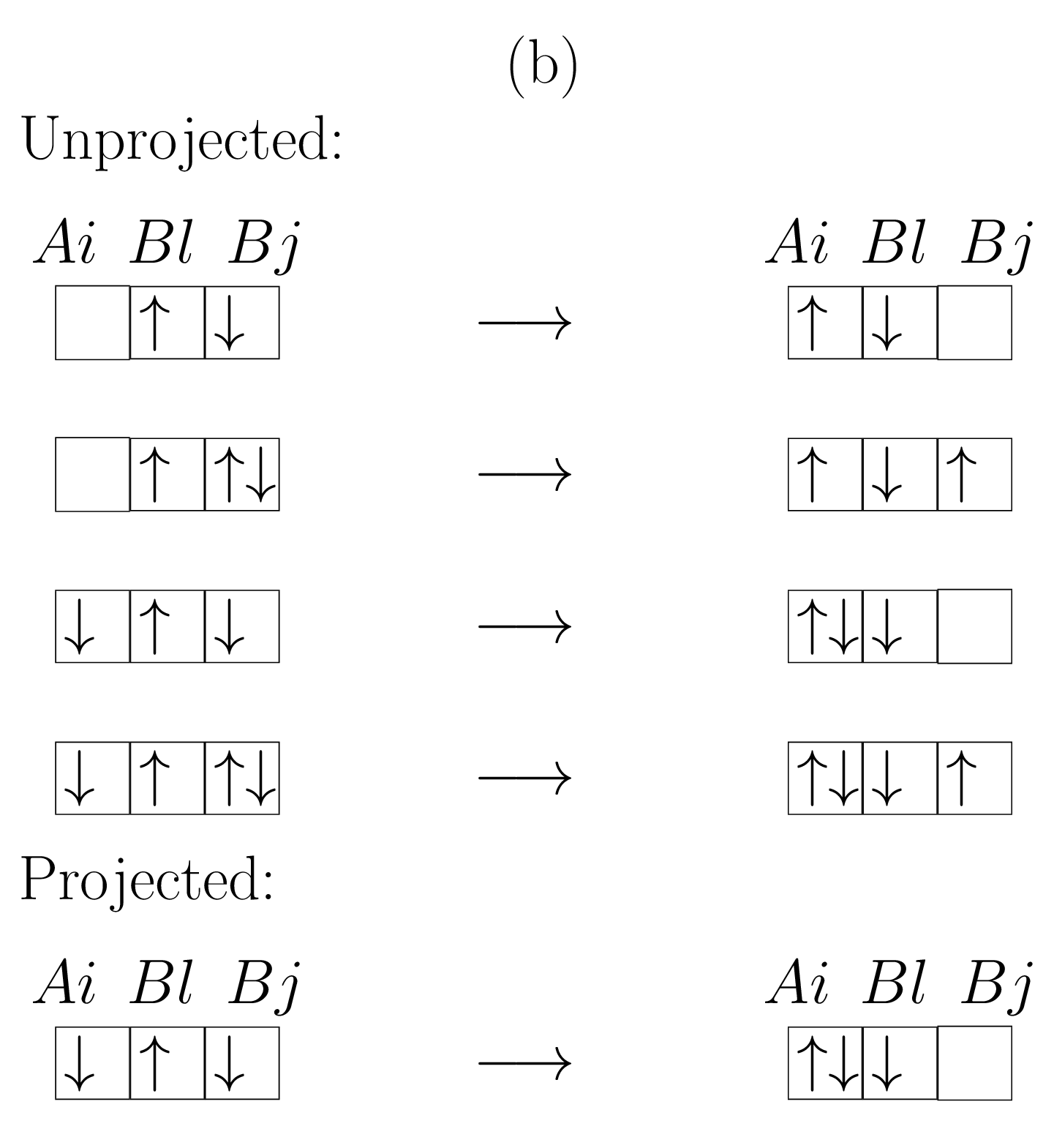}}}
\hfill
\caption{(a) Processes involved in the calculation of  $g_{1\sigma}^{BBA}$. Similar BBA physical processes with doublon at intermediate B site in the unprojected basis are considered in the calculation. (b) Processes involved in calculation of $g_{2\sigma}^{BBA}$.}
\label{BBAGutz}
\end{figure*}

Now we will focus on the Gutzwiller factors of the new trimer terms which arise out of  AAB and BBA processes. The renormalised $AAB$ and $BBA$ trimer terms can be expressed as 
\bea
H_{trimer}^{A_i,A_k,B_j}=K\sum_{\sigma}(g_{1\sigma}^{AAB}(i,k,j)c_{iA\sigma}^{\dagger}(1-n_{kA\bar{\sigma}})c_{jB\sigma} \nonumber \\
+g_{2\sigma}^{AAB}(i,k,j) c_{iA\sigma}^{\dagger}c_{kA\bar{\sigma}}^{\dagger}c_{kA\sigma}c_{jB\bar{\sigma}}) \nonumber \\
H_{trimer}^{B_j,B_l,A_i}=-K\sum_{\sigma}(g_{1\sigma}^{BBA}(j,l,i)c_{iA\sigma}^{\dagger}n_{lB\bar{\sigma}}c_{jB\sigma} \nonumber \\
-g_{2\sigma}^{BBA}(j,l,i) c_{iA\sigma}^{\dagger}c_{lB\bar{\sigma}}^{\dagger}c_{lB\sigma}c_{jB\bar{\sigma}})
\eea
The AAB and BBA spin preserving hopping as depicted in Fig.~[\ref{AABGutz}(a)] and [\ref{BBAGutz}(a)] are effective next nearest neighbour hopping processes, the Gutzwiller factor for which are like nearest neighbour AB hopping. 
If we look at Fig.~[\ref{AABGutz}(a)] for the processes involved in the calculation of the Gutzwiller factor $g_{1\uparrow}^{AAB}$, we see that the probability of the process in the unprojected basis is $(1-{\bf{n}}_{iA\uparrow}){\bf{n}}_{iA\uparrow}(1-{\bf{n}}_{kA\downarrow})^{2}(1-{\bf{n}}_{jB\uparrow}){\bf{n}}_{jB\uparrow}$ and in the projected basis it is $(1-{\bf{n}}_{iA\uparrow})({\bf{n}}_{iA}-1)(1-{\bf{n}}_{kA\downarrow})^{2}{\bf{n}}_{jB\uparrow}(1-{\bf{n}}_{jB})$ resulting in the Gutzwiller factor
 \be
g_{1\uparrow}^{AAB}(i,k,j)=\sqrt{\dfrac{({\bf{n}}_{iA}-1)(1-{\bf{n}}_{jB})}{{\bf{n}}_{iA\uparrow}(1-{\bf{n}}_{jB\uparrow})}}
\ee
It is to be remembered that in the unprojected basis,  processes with either up or hole at the intermediate A site have been considered. In Fig.~[\ref{BBAGutz}(a)], processes involved in the calculation of $g_{1\uparrow}^{BBA}$ has been depicted. The probability of the process in the unprojected basis is $(1-{\bf{n}}_{iA\uparrow}){\bf{n}}_{iA\uparrow}{\bf{n}}_{lB\downarrow}^{2}(1-{\bf{n}}_{jB\uparrow}){\bf{n}}_{jB\uparrow}$ and in the projected basis is $(1-{\bf{n}}_{iA\uparrow})({\bf{n}}_{iA}-1){\bf{n}}_{lB\downarrow}^{2}(1-{\bf{n}}_{jB})n_{jB\uparrow}$. Then, the Gutzwiller factor is 
\be
g_{1\uparrow}^{BBA}(j,l,i)=\sqrt{\dfrac{({\bf{n}}_{iA}-1)(1-{\bf{n}}_{jB})}{{\bf{n}}_{iA\uparrow}(1-{\bf{n}}_{jB\uparrow})}}
\ee
 which is the same as $g_{1\uparrow}^{AAB}(i,k,j)$.

The Gutzwiller factors for spin flip terms depicted in Fig.~[\ref{AABGutz}(b)] and [\ref{BBAGutz}(b)] can be found out similarly. For $g_{2\uparrow}^{AAB}(i,k,j)$, the probability in the unprojected space is $(1-{\bf{n}}_{iA\uparrow}){\bf{n}}_{iA\uparrow}(1-{\bf{n}}_{kA\uparrow})(1-{\bf{n}}_{kA\downarrow}){\bf{n}}_{kA\uparrow}{\bf{n}}_{kA\downarrow}{\bf{n}}_{jB\downarrow}(1-{\bf{n}}_{jB\downarrow})$ and in the projected space is $(1-{\bf{n}}_{iA\uparrow})({\bf{n}}_{iA}-1)(1-{\bf{n}}_{kA\uparrow})(1-{\bf{n}}_{kA\downarrow}){\bf{n}}_{jB\downarrow}(1-{\bf{n}}_{jB})$ resulting in the Gutzwiller factor
\be
g_{2\uparrow}^{AAB}(i,k,j)=\sqrt{\dfrac{({\bf{n}}_{iA}-1)(1-{\bf{n}}_{jB})}{{\bf{n}}_{kA\uparrow}{\bf{n}}_{kA\downarrow}{\bf{n}}_{iA\uparrow}(1-{\bf{n}}_{jB\downarrow})}}
\ee
For $g_{2\uparrow}^{BBA}(j,l,i)$, the probability in the unprojected space is $(1-{\bf{n}}_{iA\uparrow}){\bf{n}}_{iA\uparrow}{\bf{n}}_{lB\uparrow}{\bf{n}}_{lB\downarrow}(1-{\bf{n}}_{lB\uparrow})(1-{\bf{n}}_{lB\downarrow}){\bf{n}}_{jB\downarrow}(1-{\bf{n}}_{jB\downarrow})$ and that in the projected space is $(1-{\bf{n}}_{iA\uparrow})({\bf{n}}_{iA}-1){\bf{n}}_{lB\uparrow}{\bf{n}}_{lB\downarrow}{\bf{n}}_{jB\downarrow}(1-{\bf{n}}_{jB})$ leading to the Gutzwiller factor
 \be
g_{2\uparrow}^{BBA}(j,l,i)=\sqrt{\dfrac{({\bf{n}}_{iA}-1)(1-{\bf{n}}_{jB})}{(1-{\bf{n}}_{lB\uparrow})(1-{\bf{n}}_{lB\downarrow}){\bf{n}}_{iA\uparrow}(1-{\bf{n}}_{jB\downarrow})}}
\ee

\subsection{Insights into correlated binary alloy from the renormalised Hamiltonian}
The renormalised Hamiltonian derived above brings deep insight towards the possible phase diagram of the strongly correlated binary alloy. Let us first focus at the projected hopping terms given in Eq.~(\ref{H1b}) and the corresponding Gutzwiller factors in Eq.~(\ref{gtb}). At half-filling for $U \gg t$, if the disorder is weak, system will be an antiferromagnetic Mott insulator because the hopping  term is completely projected out. As disorder increases and becomes comparable to $U$, the local particle density does not remain close to one on all the sites and the Gutzwiller factors $g_{t\sigma}^{\alpha\beta}$ for various hopping processes become finite resulting in finite kinetic energy of the electrons. Also the Mott gap reduces with increase in $V$. This indicates towards the possibility of a metallic phase in the system for $V \sim U$. This is consistent with what has been shown within DMFT + coherent potential approximation~\cite{Potthoff}. In the metallic phase, the quasiparticle weight will be given by the most probable value of the Gutzwiller factors for hopping terms (in Eq.~(\ref{gtb})). Since $V \sim U$, the local electron densities will not deviate much from unity. Hence the Gutzwiller factors $g_{t\sigma}^{\alpha\beta}$ are very small resulting in very small quasiparticle weight in the metallic phase. 

Let us now turn our attention to the spin exchange terms in the low energy Hamiltonian. For the parameter regime $V \sim U \gg t$, since the effective hopping in the projected Hilbert space becomes finite, and the electron density on each site is not one, spin exchange terms might give rise to disordered superconductivity with either $d$ wave pairing or $d+is$ pairing. Due to the presence of large binary disorder, we might get disordered superconducting phase coexisting with an incommensurate/dis-commensurate charge density wave which is a topic of great interest in context of high $T_c$ superconductors~\cite{cdwsc}.  
\section{Conclusion}
In this work, we have extended the idea of Gutzwiller projection for excluding holes from the low energy Hilbert space, which so far has been developed only for exclusion of doublons e.g. in context of hole doped $t-J$ model. We have discussed variants of Hubbard model with onsite potentials because of which in the limit of strong correlations and comparable potential terms, on some sites doublons are projected out from low energy Hilbert space while from some other sites holes are projected out from the low energy Hilbert space. In order to understand the physics of these systems, it becomes essential to understand how to carry out Gutzwiller projection for holes. We defined new fermionic operators in case of hole projected Hilbert space and derived effective low energy Hamiltonian for these models by carrying out systematic similarity transformation. We further carried out rescaling of couplings in the effective Hamiltonian using Gutzwiller approximation to implement the effect of site dependent projection of holes and doublons. To be specific, we provided details of similarity transformation and Gutzwiller approximation for IHM and Hubbard model with binary disorder. 

The effective low energy Hamiltonian derived in both the cases shines light into the possibility of exotic phases. In the half filled IHM, our renormalised Hamiltonian predicts a half-metal phase followed up by a metal with increase in $\Delta$ for $U \sim \Delta$ and a superconducting phase for higher dimensional ($d\ge 2$) systems. Our effective Hamiltonian also explains the  non-monotonic behaviour of the Neel temperature as a function of $\Delta$ in the AFM phase of the IHM realised for $U \gg t$. 
In the correlated binary alloy model, for both disorder and e-e interactions being much larger than the hopping amplitude ($V\sim U \gg t$), there is a possible metallic phase which might turn into a very narrow disordered superconducting phase coexisting with dis-commensurate charge density wave in two or higher dimensional systems with the help of effective next nearest neighbour hopping. The nature of Gutzwiller factors indicate that the metallic phase intervening the two insulating phases in the IHM or the correlated binary alloy model will be a bad metal with very high effective mass of the quasiparticles.

Although we have considered so far case of strongly correlated Hubbard model in the presence of large binary disorder, the formalism can be easily used even in case of fully random disorder $V(i)\in [-V,V]$.  Strongly correlated Hubbard model in the presence of fully random disorder has been mostly studied in the limit of weak disorder mainly in context of high $T_c$ cuprates~\cite{garg_nature,AG_NP}. Case of strong disorder has been studied in order to understand the effect of impurities like $Zn$ in high $T_c$ cuprates ~\cite{Zn} but that too keeping $V \le U$ so that the constraint of no double occupancy remains intact. But for the limit of strong correlation as well as strong disorder such that $U \sim V \gg t$ the formalism of hole projection is essential and has not been studied before. For $V(i) <0$ and $|V(i)|> V_c$, where $V_c \gg t$, holes will not be allowed in the low energy Hilbert space. Though due to the limit of strong correlations for hole-doped case, doublons will not be energetically allowed at other sites of the system which have either $V(i) > 0$ or $ V(i) < 0$ but $|V(i)| < V_c$. Hence, even in case of fully random disorder there will be effectively two type of sites $A$ where holes are  projected out from low energy Hilbert space and $B$ type sites where doublons are projected out from low energy Hilbert space and one can easily use the formalism we have provided for strongly correlated binary alloys. Another situation where this physics is of relevance is a strongly correlated Hubbard model with large attractive impurities randomly distributed over the lattice with $V(i) =-V$ at the impurity sites and $V(i)=0$ at other sites of the lattice. For $V \sim U \gg t$, at the impurity sites energetics will not allow holes in the low energy Hilbert space while at all other sites of the lattice for which $V(i)=0$ large $U$ will not allow for doublons in the low energy sector for hole-doped case. Again in this situation one can use the formalism  developed here for the case of strongly correlated binary alloys. 
 
To conclude, in this work we have provided an essential tool which has been missing so far in the field of stongly correlated electron systems, that is, the Gutzwiller projection for holes allowing for doublons which happens in many correlated systems in various possible scenarios explained above. We have descibed its implementation at the level of Gutzwiller approximation. We would like to mention that so far we have evaluated Gutzwiller factors under simplest assumption of spin resolved densities being same in the projected and unprojected state. In future work we would like to extend this work to find Gutzwiller factors in more general scenarios.


\begin{thebibliography}{9}
\bibitem{footnote}
Note that there is a difference in convention used for $\Delta$ term in the Hamiltonian in ~\cite{Soumen} and current Hamiltonian. The two are related by a factor of 2)
\end{thebibliography}
\end{document}